\documentclass[english,aps,prc,superscriptaddress,nofootinbib,longbibliography]{revtex4-1}
\usepackage[T1]{fontenc}
\usepackage[latin9]{inputenc}
\usepackage{float}
\usepackage{amsmath}
\usepackage{amssymb}
\usepackage{graphicx}
\usepackage{esint}

\makeatletter

\usepackage{babel}

\usepackage{babel}

\makeatother

\usepackage{babel}
\begin{document}

\title{Local spin polarization in high energy heavy ion collisions}

\author{Hong-Zhong Wu}

\email{whz168@mail.ustc.edu.cn}

\selectlanguage{english}%

\affiliation{Department of Modern Physics, University of Science and Technology
of China, Hefei, Anhui 230026, China}

\author{Long-Gang Pang}

\email{lgpang@lbl.gov}

\selectlanguage{english}%

\affiliation{Nuclear Science Division, MS 70R0319, Lawrence Berkeley National
Laboratory, Berkeley, California 94720}

\author{Xu-Guang Huang}

\email{huangxuguang@fudan.edu.cn}

\selectlanguage{english}%

\affiliation{Physics Department and Center for Particle Physics and Field Theory,
Fudan University, Shanghai 200433, China}

\affiliation{Key Laboratory of Nuclear Physics and Ion-beam Application (MOE),
Fudan University, Shanghai 200433, China}

\author{Qun Wang}

\email{qunwang@ustc.edu.cn}

\selectlanguage{english}%

\affiliation{Department of Modern Physics, University of Science and Technology
of China, Hefei, Anhui 230026, China}
\begin{abstract}
We revisit the azimuthal angle dependence of the local spin polarization
of hyperons in heavy-ion collisions at 200 GeV in the framework of
the (3+1)D viscous hydrodynamic model CLVisc. Two different initial
conditions are considered in our simulation: the optical Glauber initial
condition without initial orbital angular momentum and the AMPT initial
condition with an initial orbital angular momentum. We find that the
azimuthal angle dependence of the hyperon polarization strongly depends
on the choice of the so-called \textit{spin chemical potential} $\Omega_{\mu\nu}$.
With $\Omega_{\mu\nu}$ chosen to be proportional to the temperature
vorticity, our simulation shows qualitatively coincidental results with the recent
measurements at RHIC for both the longitudinal and transverse polarization.
We argue that such a coincidence may be related to the fact that
the temperature vorticity is approximately conserved in the hot quark-gluon
matter.
\end{abstract}
\maketitle

\section{Introduction}

It is well known that the rotation and spin polarization are correlated
and can be converted to each other in materials \cite{dehaas:1915,Barnett:1935}.
Recently, the polarization of electrons in a vortical fluid has been
observed \cite{Takahashi:2016}. Similar phenomena also exist in high-energy
heavy-ion collisions in which huge orbital angular momenta (OAM) are
produced in peripheral collisions \cite{Liang:2004ph,Becattini:2007sr,Gao2008,Huang:2011ru,Wang:2017jpl}
(for a recent review, see, e.g. \cite{Wang:2017jpl}). The huge OAM
are distributed into the quark gluon plasma created in heavy-ion collisions
in the form of local vorticity \cite{Becattini:2015ska,Pang:2016igs,Deng:2016gyh,Jiang:2016woz},
which result in the local polarization of hadrons along the vorticity
direction \cite{Becattini:2013fla,Fang:2016vpj} due to the spin-orbit
coupling \cite{Liang:2004ph,Gao2008}. The net effect of the local
polarization at all space-time points on the freeze-out hyper-surface
gives the global polarization in the direction of the reaction plane
or the OAM of two colliding nuclei \cite{Liang:2004ph,Liang:2004xn,Voloshin:2004ha,Betz2007,Becattini:2007sr,Gao2008,Wang:2017jpl}.

The global polarization of $\Lambda$ and $\bar{\Lambda}$ has been
measured by the STAR collaboration in Au+Au collisions at $\sqrt{s_{NN}}=7.7-200$
GeV \cite{STAR:2017ckg,Adam:2018ivw}. The data show a decreasing
trend in collision energies from about 2\% at 7.7 GeV to about 0.3\%
at 200 GeV.

There are several theoretical approaches which have been developed
to study the global and local polarization in heavy ion collisions.
The statistic-hydro model is based on the spin-vorticity coupling
in the thermal distribution function \cite{Becattini:2013fla,Becattini:2015nva,Becattini:2016gvu,Florkowski:2017dyn,Florkowski:2018ahw}.
So the average spin polarization is proportional to the so-called
thermal vorticity (see the definition in the next section) if the thermal vorticity
is small. Another theoretical approach is the Wigner function
(WF) formalism \cite{Heinz:1983nx,Elze:1986qd,Vasak:1987um,Zhuang:1995pd,Florkowski:1995ei,Blaizot:2001nr,Wang:2001dm},
which has been revived \cite{Gao:2012ix,Chen:2012ca,Gao:2015zka,Hidaka:2016yjf,Gao:2017gfq,Gao:2018wmr,Huang:2018wdl,Gao:2018jsi,Liu:2018xip}
to study the chiral magnetic effect (CME) \cite{Vilenkin:1980fu,Kharzeev:2007jp,Fukushima:2008xe}
(for reviews, see, e.g., Ref. \cite{Kharzeev:2013jha,Kharzeev:2015znc,Huang:2015oca,Hattori:2016emy})
and chiral vortical effect (CVE) \cite{Vilenkin:1978hb,Erdmenger:2008rm,Banerjee:2008th,Son:2009tf,Gao:2012ix,Hou:2012xg}
for massless fermions. Recently, the kinetic theory for spin-1/2 massive
fermions has been formulated in the WF framework \cite{Fang:2016vpj,Weickgenannt:2019dks,Gao:2019znl,Hattori:2019ahi,Wang:2019moi},
which is useful in describing the evolution of the spin polarization.
This is because the axial vector component gives the spin phase space
distribution of fermions. At equilibrium, when the thermal vorticity
is small, the spin polarization of fermions from the WF formalism
is proportional to the thermal vorticity, consistent with the statistic-hydro
model.

To describe the STAR data on the global polarization of hyperons which
is along the direction of the reaction plane, the hydrodynamic and
transport models have been used to calculate the vorticity field \cite{Baznat:2013zx,Csernai:2013bqa,Csernai:2014ywa,Teryaev:2015gxa,Jiang:2016woz,Deng:2016gyh,Ivanov:2017dff,Li:2017slc,Wei:2018zfb}.
In the hydrodynamic framework, the velocity and in turn the vorticity
fields in the fireball can be obtained naturally. The transport
models describe the phase space evolution of a particle system through
collisions among particles, so the position and momentum of each particle
in the system at any time is given. To obtain the fluid velocity and
then the vorticity at one space-time point, the suitable coarse graining
procedure has to be used. Once the vorticity field is obtained, the
global polarization of hyperons can be calculated from an integral
over the freeze-out hyper-surface which agrees well with the data \cite{Karpenko:2016jyx,Xie:2017upb,Li:2017slc,Sun:2017xhx,Wei:2018zfb}.

The polarization of hyperons as a function of the azimuthal angle
in the transverse plane has been recently measured in the STAR experiment
\cite{Adam:2018ivw,Adam:2019srw}. However the data for the polarization
along both the longitudinal and the transverse directions cannot be
described by the hydrodynamic models (including A Multi-phase Transport
(AMPT) model from which the vorticity field is extracted by the coarse
graining method) \cite{Becattini:2017gcx,Xia:2018tes,Wei:2018zfb}
based on the coupling of the thermal vorticity and the spin at equilibrium.
There is a sign difference between the data and these model calculations.
Although one model based on the chiral kinetic theory can explain
the sign of the data \cite{Sun:2018bjl}, it cannot reproduce the
magnitude of the data. Recent studies showed that the feed-down effects
cannot resolve the sign difference \cite{Xia:2019fjf,Becattini:2019ntv}.

The disagreement between theories and experiments indicates that the
spin degree of freedom may not reach equilibrium in the fireball and
thus the spin polarization may not be determined by the thermal vorticity.
The relation between spin and thermal vorticity is dictated
by the condition of local thermodynamic equilibrium if the
spin tensor does not play a physical role \cite{Becattini:2018duy}.
This calls for new approaches, for examples, the spin can be treated as an independent
dynamical variable in the spin kinetic theory and spin hydrodynamics, or
dissipative terms should be considered which are possibly larger than believed.
Recently, the framework of spin hydrodynamics was developed~\cite{Florkowski:2017ruc,Florkowski:2018fap,Hattori:2019lfp}.
The spin evolution based on particle collisions was derived \cite{Zhang:2019xya}.
The purpose of the present paper is not to make a numerical study
based on these new approaches, instead, our purpose is not that ambitious:
we will explore different choices of the so-called ``spin chemical
potential'' $\Omega_{\mu\nu}$ and calculate the corresponding local
hyperon polarization. The underlying reason is that, beyond global
equilibrium, the thermal vorticity is not guaranteed to be the spin
chemical potential, and thus the latter becomes a free parameter~\cite{Florkowski:2018fap,Hattori:2019lfp,Becattini:2018duy}.
In the (3+1)D hydrodynamic model CLVisc \cite{Pang:2012he,Pang:2018zzo},
we will assume that the spin chemical potential $\Omega_{\mu\nu}$
is still determined by the fluid velocity and temperature (or equivalently
the energy density). This means that $\Omega_{\mu\nu}$, being an
anti-symmetric tensor, can be regarded as a type of vorticity (with
appropriate normalization to make the dimension correct). We will
thus explore four different definitions for $\Omega_{\mu\nu}$ or
vorticity and calculate the local hyperon polarization and compare
with the data. In our hydrodynamic simulation, we will examine two
different initial conditions: the optical Glauber initial condition
without initial OAM and the AMPT initial condition with an initial
OAM.

The paper is organized as follows. In Section II we give a brief discussion
about our motivation. In Section III we introduce our hydrodynamic
model which we use for the simulation. We present our numerical results
in Section IV. We give some discussions in Section V. Finally, we give a summary of our results in Section
VI.

\section{Spin polarization and vorticity}

The thermodynamic equilibrium in quantum field theory can be described
by the density operator $\hat{\rho}$. Its form at local equilibrium
can be obtained by maximizing the entropy $S=-\mathrm{Tr}(\hat{\rho}\ln\hat{\rho})$
with fixed densities of the energy-momentum, the angular momentum,
and the conserved charge current on a space-like hyper-surface $\Sigma^{\mu}=n^{\mu}\Sigma$
pointing to a time-like direction $n^{\mu}$ \cite{zubarev:1979,weert:1982,Becattini:2014yxa,Hayata:2015lga},
\begin{eqnarray}
n_{\mu}\mathrm{Tr}(\hat{\rho}\hat{T}^{\mu\nu}) & = & n_{\mu}T^{\mu\nu},\nonumber \\
n_{\mu}\mathrm{Tr}(\hat{\rho}\hat{J}^{\mu,\alpha\beta}) & = & n_{\mu}J^{\mu,\alpha\beta},\nonumber \\
n_{\mu}\mathrm{Tr}(\hat{\rho}\hat{N}^{\mu}) & = & n_{\mu}N^{\mu},\label{eq:density-op}
\end{eqnarray}
where $\hat{T}^{\mu\nu}$, $\hat{J}^{\mu,\alpha\beta}$ and $\hat{N}^{\mu}$
are the density operators of the energy-momentum tensor, the angular
momentum tensor, and the conserved charge current, respectively. Note
that $\hat{T}^{\mu\nu}$ is not necessarily symmetric. The quantities
$T^{\mu\nu}$, $J^{\mu,\alpha\beta}$ and $N^{\mu}$ are their expectation
values. For simplicity we will call $\hat{T}^{\mu\nu}$ ($T^{\mu\nu}$)
and $\hat{J}^{\mu,\alpha\beta}$ ($J^{\mu,\alpha\beta}$) the energy-momentum
and angular momentum tensor respectively though they are actually
tensor densities. The angular momentum density operator includes the
orbital and spin parts
\begin{equation}
\hat{J}^{\mu,\alpha\beta}=x^{\alpha}\hat{T}^{\mu\beta}-x^{\beta}\hat{T}^{\mu\alpha}+\hat{S}^{\mu,\alpha\beta}.\label{eq:angular-momentum}
\end{equation}
Thus, the second constraint in Eq. (\ref{eq:density-op}) can be equivalently
expressed as
\begin{equation}
n_{\mu}\mathrm{Tr}(\hat{\rho}\hat{S}^{\mu,\alpha\beta})=n_{\mu}S^{\mu,\alpha\beta}.\label{new_constraint on angular momentum}
\end{equation}

The form of the density operator under the constraints (\ref{eq:density-op}),
or with the second constraint in Eq. (\ref{eq:density-op}) being
replaced by the constraint (\ref{new_constraint on angular momentum}),
that maximizes the entropy reads
\begin{equation}
\hat{\rho}_{{\rm LE}}=\frac{1}{Z_{{\rm LE}}}\mathrm{exp}\left[-\int d\Sigma_{\mu}\left(\hat{T}^{\mu\nu}\beta_{\nu}-\frac{1}{2}\Omega_{\alpha\beta}\hat{S}^{\mu,\alpha\beta}-\zeta\hat{N}^{\mu}\right)\right],\label{eq:density_operator}
\end{equation}
where $\beta_{\nu}$, $\Omega_{\alpha\beta}$ and $\zeta$ are Lagrangian
multipliers which have physical meanings: $\beta_{\nu}=u_{\nu}/T$
with $u_{\nu}$ being the four-velocity and $T$ being the temperature,
$\zeta=\mu/T$ with $\mu$ being the chemical potential, and $\Omega_{\alpha\beta}$
plays the role of the chemical potential for the angular momentum~\footnote{More precisely, it is $T\Omega_{\alpha\beta}$ that plays the role
of a chemical potential for the angular momentum.}. In the following, we will simply call $\Omega_{\alpha\beta}$ the
spin chemical potential as it determines the spin polarization at
local equilibrium. The density operator $\hat{\rho}_{{\rm LE}}$ in
(\ref{eq:density_operator}) defines the local thermal equilibrium
and in general depends on the time.

In relativistic hydrodynamics, in order to obtain the spin vector,
we need to first obtain $T$, $u^{\mu}$, and $\Omega_{\mu\nu}$ by solving
the hydrodynamic equations in which the spin degree of freedom (or
equivalently $\Omega_{\mu\nu}$) is treated on the same footing as
$T$ and $u^{\mu}$. Such a framework is the spin hydrodynamics~\cite{Florkowski:2017ruc,Hattori:2019lfp}.
However, the numerical spin hydrodynamics has not been established
yet. Therefore we will adopt an usual (3+1)D hydrodynamic model, CLVisc
\cite{Pang:2012he,Pang:2018zzo}, which can give the space-time evolution
of $T$ and $u^{\mu}$. Since $\Omega_{\mu\nu}$ is antisymmetric,
we then assume that $\Omega_{\mu\nu}$ can be constructed from $T$
and $u^{\mu}$ as $\Omega_{\mu\nu}=-(1/2)\lambda(T)[\partial_{\mu}(g(T)u_{\nu})-\partial_{\nu}(g(T)u_{\mu})]\equiv\lambda(T)\omega_{\mu\nu}$
or its projections where $\lambda$ and $g$ are scalar functions
of $T$ and $\omega_{\mu\nu}$ is the vorticity tensor~\footnote{In principle, it is also allowed to use the Hodge dual of the vorticity
tensor to construct $\Omega_{\mu\nu}$. However, when the global equilibrium
is approached, it is known that $\Omega_{\mu\nu}$ should approach
the thermal vorticity up to a constant (depending on the symmetry
properties of $T^{\mu\nu}$). We therefore do not consider such a
possibility here.}. In our numerical simulation, four types of vorticity will be considered,
namely, the kinematic vorticity, the relativistic extension of the
non-relativistic vorticity (NR vorticity), the thermal vorticity,
and the temperature vorticity (T-vorticity). 
All these four types of vorticities have definite physical meaning 
and have been widely studied for quite some time in literature. 
For example, in Ref. \cite{Becattini:2015ska} 
the properties of the kinematic, thermal and T-vorticity have been 
studied in heavy ion collisions, in Ref. \cite{Karpenko:2018erl} the NR vorticity 
is defined in a study of Lambda polarization in heavy ion collisions.

The kinematic vorticity is defined by
\begin{equation}
\omega_{\mu\nu}^{(K)}=-\frac{1}{2}(\partial_{\mu}u_{\nu}-\partial_{\nu}u_{\mu}),\label{kinematical-vorticity}
\end{equation}
where $u^{\mu}=\gamma(1,\boldsymbol{v})$ is the four-velocity and
$\gamma$ is the Lorentz factor. The mass dimension of the kinematic
vorticity is 1. The kinematic vorticity tensor can be decomposed
into the part parallel and the part orthogonal to the fluid velocity
\begin{equation}
\omega_{\mu\nu}^{(K)}=\varepsilon_{\nu}u_{\mu}-\varepsilon_{\mu}u_{\nu}+\epsilon_{\nu\mu\rho\eta}u^{\rho}\omega^{\eta},\label{eq:decomp-k-vort}
\end{equation}
where $\varepsilon_{\mu}=-(1/2)u^{\nu}\partial_{\nu}u_{\mu}$ and
$\omega^{\mu}=(1/2)\epsilon^{\mu\nu\rho\sigma}u_{\nu}(\partial_{\rho}u_{\sigma})$.
In comparison with the decomposition of the electromagnetic field
strength $F^{\mu\nu}$, the vector $\varepsilon^{\mu}$ is like an
`electric' field while the vorticity vector $\omega^{\mu}$ is like
a `magnetic' field. It is clear that $\omega^{\mu}$ is a direct extension
of the vorticity defined in non-relativistic hydrodynamics, $\boldsymbol{\omega}=(1/2)\nabla\times\boldsymbol{v}$.
We thus define the last term in Eq. (\ref{eq:decomp-k-vort}) as the
NR vorticity tensor \cite{Karpenko:2018erl}
\begin{equation}
\omega_{\mu\nu}^{(\mathrm{NR})}=\epsilon_{\nu\mu\rho\eta}u^{\rho}\omega^{\eta}.\label{eq:non-rel}
\end{equation}
Similar to $\omega_{\mu\nu}^{(\mathrm{NR})}$, there has been an attempt
to use the spatial components of the thermal vorticity as the spin
chemical potential to study the longitudinal spin polarization of
hyperons~\cite{Florkowski:2019voj}.

The temperature vorticity or T-vorticity is defined by
\begin{align}
\omega_{\mu\nu}^{(T)} & =-\frac{1}{2}[\partial_{\mu}(Tu_{\nu})-\partial_{\nu}(Tu_{\mu})]\nonumber \\
 & =T\omega_{\mu\nu}^{(K)}+\frac{1}{2}(u_{\mu}\partial_{\nu}T-u_{\nu}\partial_{\mu}T)\nonumber \\
 & \equiv T\omega_{\mu\nu}^{(K)}+\omega_{\mu\nu}^{(T)}(T),\label{T vorticity}
\end{align}
where the temperature enters the space-time derivative. The mass dimension
of the temperature vorticity is 2. We have decomposed $\omega_{\mu\nu}^{(T)}$
into the part involving the space-time gradient of the temperature
and the part without it which is proportional to the kinematic vorticity.

An important property of the T-vorticity is that it obeys a conservation
law \cite{Gao:2014coa,Becattini:2015ska,Deng:2016gyh}. Suppose $\Xi$
is a two-dimensional hyper-surface and $C$ is its boundary, thus the
flux of the temperature vorticity on $\Xi$ is equal to the corresponding
circulation of $Tu^{\mu}$ along the boundary $C$
\begin{equation}
\int_{\Xi}\omega_{\mu\nu}^{(T)}dx^{\mu}\wedge dx^{\nu}=-\oint_{C}Tu_{\mu}dx^{\mu}.
\end{equation}
Since the viscosity of the hot matter in the fireball is small, we
can approximately apply the Euler equation for an ideal fluid
\begin{equation}
(\varepsilon+P)\frac{d}{d\tau}u^{\mu}=\nabla^{\mu}P,\label{eq:releuler}
\end{equation}
where $\varepsilon$ and $P$ are the energy density and pressure
respectively, $d/d\tau=u^{\mu}\partial_{\mu}$ is co-moving time derivative,
and $\nabla_{\mu}=\partial_{\mu}-u_{\mu}(d/d\tau)$. Rewriting the
Euler equation in the following form
\begin{equation}
\frac{d}{d\tau}(Tu^{\mu})=\partial^{\mu}T,
\label{eq:euler2}
\end{equation}
 one easily finds
\begin{equation}
\frac{d}{d\tau}\oint Tu_{\mu}dx^{\mu}=\oint\partial_{\mu}Tdx^{\mu}=0.
\label{eq:cons-tu}
\end{equation}
This is the relativistic Helmholtz-Kelvin theorem: the flux of the
T-vorticity tensor is conserved with the fluid cell along $u^{\mu}$.
We will see that this imposes a strong influence on the spin polarization.

The thermal vorticity $\omega_{\mu\nu}^{(\mathrm{th})}$ is defined
by
\begin{align}
\omega_{\mu\nu}^{(\mathrm{th})} & =-\frac{1}{2}[\partial_{\mu}(\beta u_{\nu})-\partial_{\nu}(\beta u_{\mu})]\nonumber \\
 & =\frac{1}{T}\omega_{\mu\nu}^{(K)}-\frac{1}{2T^{2}}(u_{\mu}\partial_{\nu}T-u_{\nu}\partial_{\mu}T)\nonumber \\
 & =\frac{1}{T}\omega_{\mu\nu}^{(K)}+\omega_{\mu\nu}^{(\mathrm{th})}(T),\label{thermal vorticity}
\end{align}
where $\beta=1/T$. The thermal vorticity is dimensionless. Similar
to $\omega_{\mu\nu}^{(T)}$ we have also decomposed $\omega_{\mu\nu}^{(\mathrm{th})}$
into the part involving the space-time gradient of the temperature
and the part without it which is proportional to $\omega_{\mu\nu}^{(K)}$.
We see in Eq. (\ref{T vorticity}) and (\ref{thermal vorticity})
that $\omega_{\mu\nu}^{(T)}(T)$ and $\omega_{\mu\nu}^{(\mathrm{th})}(T)$
have the opposite sign.

The importance of the thermal vorticity relies on the fact that at
global equilibrium, $\Omega_{\mu\nu}$ equals to $\omega_{\mu\nu}^{(\mathrm{th})}$
provided that the energy-momentum tensor $\hat{T}^{\mu\nu}$ has a
non-vanishing anti-symmetric component~\cite{Becattini:2018duy,Hattori:2019lfp,LiuYC:2019}.
This can be seen from the following procedure (the analysis based
on the dissipative spin hydrodynamics or the kinetic theory gives
the same conclusion). The global equilibrium is the state that the
density operator (\ref{eq:density_operator}) becomes independent
of the choice of the hyper-surface $\Sigma_{\mu}$, so that
\begin{equation}
\hat{T}^{\mu\nu}\partial_{\mu}\beta_{\nu}-\frac{1}{2}\hat{S}^{\mu,\alpha\beta}\partial_{\mu}\Omega_{\alpha\beta}+\frac{1}{2}(\hat{T}^{\alpha\beta}-\hat{T}^{\beta\alpha})\Omega_{\alpha\beta}=0,\label{eq:independent equation}
\end{equation}
where we used $\partial_{\mu}\hat{S}^{\mu,\alpha\beta}=\hat{T}^{\beta\alpha}-\hat{T}^{\alpha\beta}$.
The above condition is fulfilled when~\footnote{We also note that these are sufficient but not necessary conditions for global equilibrium. For example, for conformal fluid, the right-hand side of the first condition can
be relaxed to $\phi(x)g_{\mu\nu}$ with $\phi$ a scalar~\cite{Liu:2018xip}.}
\begin{eqnarray}
\partial_{\mu}\beta_{\nu}+\partial_{\nu}\beta_{\mu} & = & 0,\nonumber \\
\partial_{\mu}\Omega_{\alpha\beta} & = & 0,\nonumber \\
\Omega_{\alpha\beta} & = & \omega_{\alpha\beta}^{(\mathrm{th})}.
\label{eq:conditions}
\end{eqnarray}
Note that if $\hat{T}^{\mu\nu}$ is symmetric, the spin tensor is
conserved $\partial_{\mu}\hat{S}^{\mu,\alpha\beta}=0$ and the third
condition in (\ref{eq:conditions}) does not hold which means that
$\Omega_{\mu\nu}$ remains an independent variable even at global
equilibrium. Note that the first line of Eq. (\ref{eq:conditions})
is called the Killing equation \cite{Becattini:2012tc} whose solution is
$\beta_{\mu}=b_{\mu}+\omega_{\mu\alpha}^{\mathrm{(th)}} x^{\alpha}$, where $b_{\mu}$
and $\omega_{\mu\alpha}^{\mathrm{(th)}}$ are constants.

For spin-1/2 fermions at local equilibrium, when $\omega_{\mu\nu}^{(\mathrm{th})}$
is small, the average spin vector (defined as the Pauli-Lubanski vector)
over the hyper-surface $\Sigma_{\mu}$ can be expressed as~\cite{Becattini:2013fla,Fang:2016vpj,LiuYC:2019}
\begin{equation}
S^{\mu}(p)=-\frac{1}{8m}\epsilon^{\mu\rho\sigma\tau}p_{\tau}\frac{\int d\Sigma_{\lambda}p^{\lambda}
\omega_{\rho\sigma}^{(\mathrm{th})}
f_{FD}(1-f_{FD})}{\int d\Sigma_{\lambda}p^{\lambda}f_{FD}}+O((\omega_{\mu\nu}^{(\mathrm{th})})^{2}),
\label{average-spin-vector-th}
\end{equation}
where $f_{FD}=1/[\exp(p_{\mu}\beta^{\mu}-\zeta)+1]$ is the Fermi-Dirac
distribution, normally $\Sigma_{\mu}$ is chosen as the freeze-out
hyper-surface for hyperon polarization at the freeze-out. In the calculation
we will set $\zeta=0$ as the net baryon density is almost zero in
the hot fireball created in heavy ion collisions at high energies.
In this paper, we assume that Eq. (\ref{average-spin-vector-th}) can be generalized
by replacing $\omega_{\mu\nu}^{(\mathrm{th})}$ with the spin chemical potential
$\Omega_{\mu\nu}$ as
\begin{equation}
S^{\mu}(p)=-\frac{1}{8m}\epsilon^{\mu\rho\sigma\tau}p_{\tau}\frac{\int d\Sigma_{\lambda}p^{\lambda}
\Omega_{\rho\sigma}f_{FD}(1-f_{FD})}{\int d\Sigma_{\lambda}p^{\lambda}f_{FD}}+O(\Omega_{\mu\nu}^{2}),
\label{average-spin-vector}
\end{equation}
where we will consider four types of vorticities as the spin chemical potentials, namely,
$\Omega_{\rho\sigma} = \frac{1}{T}\omega_{\rho\sigma}^{(K)}$,
$\frac{1}{T^{2}}\omega_{\rho\sigma}^{(T)}$, $\omega_{\rho\sigma}^{(\mathrm{th})}$,
$\frac{1}{T}\omega_{\rho\sigma}^{(\mathrm{NR})}$.
Here we have chosen suitable factors, $\lambda(T)=1/T,1/T^{2},1,1/T$,
respectively, to make the spin chemical potential dimensionless.
Note that Eq. (\ref{average-spin-vector}) is the main assumption of this paper.

In the following, we will use the (3+1)D hydrodynamic model CLVisc to
calculate the four types of vorticities and then use Eq. (\ref{average-spin-vector})
to obtain the spin vector and then the corresponding spin polarization.

\section{Hydrodynamic model}

The space-time evolution of the hot quark gluon plasma and dense hadronic
matter is described by second order relativistic hydrodynamic equations,
\begin{equation}
\nabla_{\mu}T^{\mu\nu}=0,\label{hydro eq}
\end{equation}
where $T^{\mu\nu}=(\varepsilon+P)u^{\mu}u^{\nu}-Pg^{\mu\nu}+\pi^{\mu\nu}$
is the energy-momentum tensor which is symmetric, $\varepsilon$ is the local energy
density in the co-moving frame of the fluid, $P=P(\varepsilon)$ is
the pressure determined by the QCD equation of state, $u^{\mu}$ is
the fluid four-velocity obeying $u_{\mu}u^{\mu}=1$, $g^{\mu\nu}=\mathrm{diag}(1,-1,-1,-1/\tau^{2})$
is the metric tensor, $\pi^{\mu\nu}$ is the shear-stress tensor whose
evolution is solved using a separate group of equations,
\begin{equation}
\pi^{\mu\nu}=\eta_{\upsilon}\sigma^{\mu\nu}-\tau_{\pi}\left[\triangle_{\alpha}^{\mu}\triangle_{\beta}^{\nu}u^{\lambda}\nabla_{\lambda}\pi^{\alpha\beta}+\frac{4}{3}\pi^{\mu\nu}\theta\right],\label{pi_mu_nu}
\end{equation}
where $\eta_{\upsilon}$ is the shear viscous coefficient, $\sigma^{\mu\nu}\equiv2\nabla^{<\mu}u^{\nu>}\equiv2\triangle^{\mu\nu\alpha\beta}\nabla_{\alpha}u_{\beta}$
is the symmetric shear tensor, $\tau_{\pi}=5\eta_{\upsilon}/(Ts)$
is the relaxation time for the shear viscosity, $\triangle^{\mu\nu}=g^{\mu\nu}-u^{\mu}u^{\nu}$
is the projection operator that makes the resulting contracted vector
orthogonal to $u^{\mu}$, $\triangle^{\mu\nu\alpha\beta}=\frac{1}{2}(\triangle^{\mu\alpha}\triangle^{\nu\beta}+\triangle^{\mu\beta}\triangle^{\nu\alpha})-\frac{1}{3}\triangle^{\mu\nu}\triangle^{\alpha\beta}$
is the double projection operator that makes the resulting contracted
tensor symmetric, traceless and orthogonal to $u^{\mu}$, $\theta\equiv\nabla_{\mu}u^{\mu}$
is the expansion rate. The operator $\nabla_{\mu}$ is the covariant
derivative operator defined as
\begin{equation}
\nabla_{\mu}\lambda^{\nu}=\partial_{\mu}\lambda^{\nu}+\Gamma_{\mu\rho}^{\nu}\lambda^{\rho},\label{eq:nabla_mn_1}
\end{equation}
\begin{equation}
\nabla_{\mu}\lambda^{\rho\sigma}=\partial_{\mu}\lambda^{\rho\sigma}+\Gamma_{\mu\lambda}^{\rho}\lambda^{\lambda\sigma}+\Gamma_{\mu\lambda}^{\sigma}
\lambda^{\rho\lambda},\label{eq:nabla_mu}
\end{equation}
for vectors $\lambda^{\mu}$ and tensors $\lambda^{\mu\nu}$. The $\Gamma$'s are Christoffel symbols
solved as a function of $g^{\mu\nu}$,
\begin{equation}
\Gamma_{\rho\sigma}^{\mu}=\frac{1}{2}g^{\mu\lambda}(\partial_{\sigma}g_{\lambda\rho}+\partial_{\rho}g_{\lambda\sigma}
-\partial_{\lambda}g_{\rho\sigma}).\label{Gamma}
\end{equation}

The above (3+1)D viscous hydrodynamic equations are solved numerically
using CLVisc \cite{Pang:2012he,Pang:2018zzo} with s95p-pce lattice
QCD equation-of-state \cite{Borsanyi:2012ve}, and two different initial
conditions: optical Glauber initial condition without initial OAM
and AMPT initial condition with initial OAM are applied to check the
dependences of the results on initial conditions.

\section{Numerical results for hyperon polarization}

\begin{figure}[H]
\begin{centering}
\includegraphics[scale=0.3]{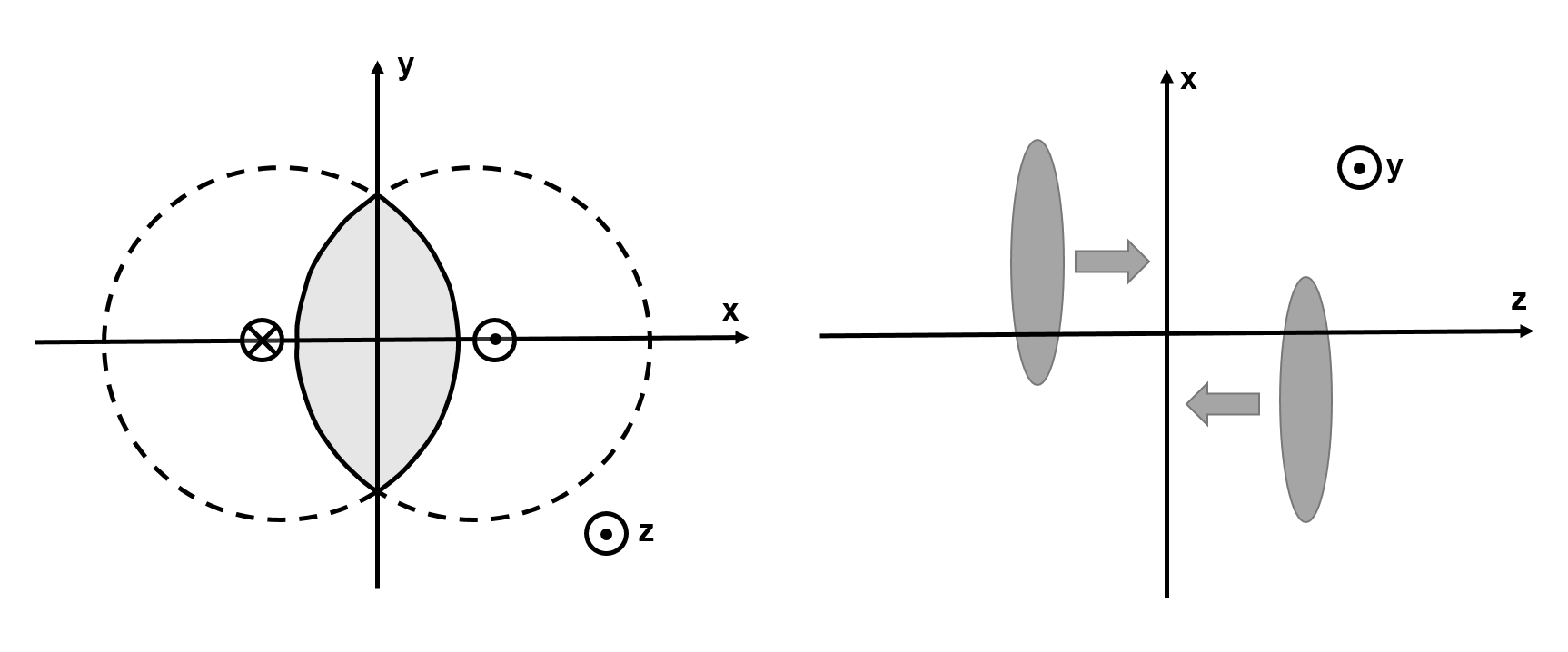}
\par\end{centering}
\caption{The coordinate system in the transverse plane. The initial OAM is
along $-y$ direction. \label{fig:collisions}}
\end{figure}

In this section we will present our numerical results for the polarization
of $\Lambda$ hyperons through vorticity fields by Eq. (\ref{average-spin-vector}).
We choose the coordinate system for collisions of two gold nuclei
at 200 GeV in 20-50\% centrality, see Fig. \ref{fig:collisions}.
The spatial indices $\mu=1,2,3$ in $S^{\mu}(p)$ in (\ref{average-spin-vector})
correspond to the $x$, $y$ and $z$ direction respectively, so sometimes
we write $\mu=1,2,3$ as $\mu=x,y,z$. To test effect of different
choices for the spin chemical potential coupled to the spin tensor,
we choose four types of vorticities: the kinematic vorticity, the
T-vorticity, the thermal vorticity and the NR vorticity. We use the
hydrodynamic model CLVisc to compute the vorticity field on the freeze-out
hyper-surface.

We use two types of the initial condition: the optical Glauber initial
condition without initial OAM and AMPT initial condition with an initial
OAM. In the optical Glauber initial condition, the initial energy
density distribution is boost invariant at mid-rapidity and is symmetric
about the y-axis. As a result, it does not provide any initial OAM.
On the other hand, the AMPT initial condition uses HIJING strings.
These strings are attached to forward and backward going participants
whose distributions are not symmetric about the y-axis in non-central
collisions. On one side of the y-z plane, the midpoints of those strings
are shifted to forward rapidity in the projectile-going direction.
On the other side of the y-z plane, the midpoints of those strings
are shifted to backward rapidity in the target-going direction. This
forward-backward asymmetry in the AMPT model introduces non-zero initial
OAM along the negative y-axis. One should keep in mind that there
is no forward-backward asymmetry if the length of strings is infinity
(before string fragmentation). In that case, all strings cover mid-rapidity.
The system would be perfectly boost invariant along the space-time
rapidity and symmetric about the y-axis. This corresponds to extreme
high energy collisions where the initial OAM disappears at mid-rapidity.
It is consistent with experimental observation that the global polarization
is stronger in lower energy collisions. For collisions at low beam
energies, the strings from the AMPT model have finite fluctuating
lengths \cite{Pang:2015zrq}. The lengths of strings are determined
by the longitudinal-momentum differences between their two end points
which are quarks and diquarks from the projectile and the target,
whose longitudinal momenta are sampled from parton distribution functions.
In this way, the lengths of strings are finite and fluctuating. This
helps to propagate the forward-backward asymmetry to left-right asymmetry
at mid-rapidity, which is responsible for the initial OAM.

We calculate the transverse and longitudinal polarization of $\Lambda$
hyperons in the rapidity range $Y\in[-\Delta Y/2,\Delta Y/2]$
\begin{eqnarray}
\mathcal{P}_{x}(p) & = & \frac{2}{\Delta Y}\int_{-\Delta Y/2}^{\Delta Y/2}dYS^{x}(p),\nonumber \\
\mathcal{P}_{y}(p) & = & \frac{2}{\Delta Y}\int_{-\Delta Y/2}^{\Delta Y/2}dYS^{y}(p),\nonumber \\
\mathcal{P}_{z}(p) & = & \frac{2}{\Delta Y}\int_{-\Delta Y/2}^{\Delta Y/2}dYS^{z}(p),\label{eq:rapidity-av}
\end{eqnarray}
where $S^{\mu}(p)$ is given by Eq. (\ref{average-spin-vector}) and
$p$ denotes the four-momentum of the $\Lambda$ hyperon,
\begin{equation}
p^{\mu}=(m_{T}\mathrm{cosh}Y,p_{x},p_{y},m_{T}\mathrm{sinh}Y),
\end{equation}
with $m_{T}=\sqrt{m_{\Lambda}^{2}+p_{x}^{2}+p_{y}^{2}}$. We choose
the rapidity range $Y\in[-1,1]$ or $\Delta Y=2$ in the calculation.

\subsection{Results with optical Glauber initial condition}

\begin{figure}[H]
\begin{centering}
\includegraphics[scale=0.3]{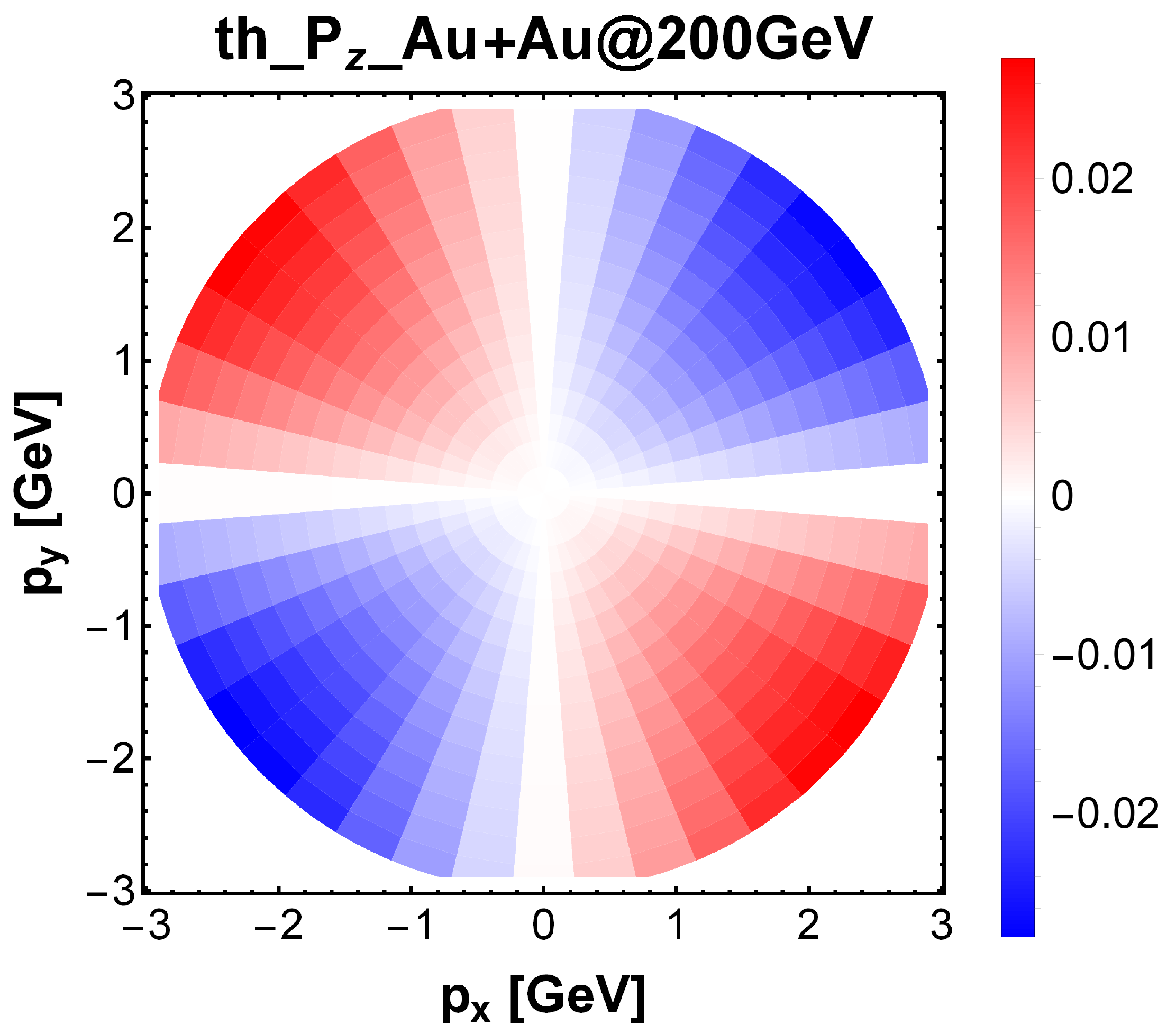}
\includegraphics[scale=0.3]{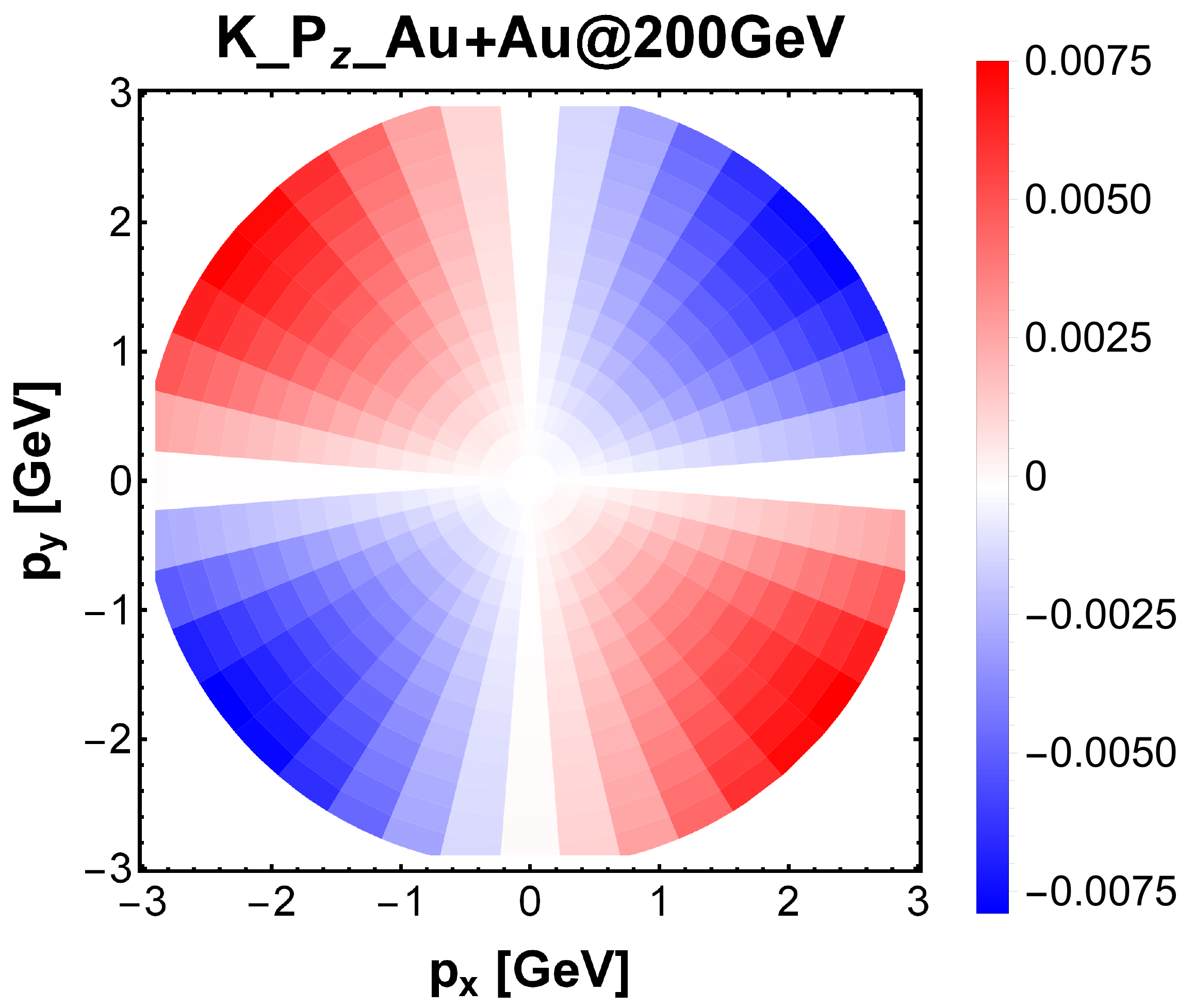}
\par\end{centering}
\centering{}
\includegraphics[scale=0.3]{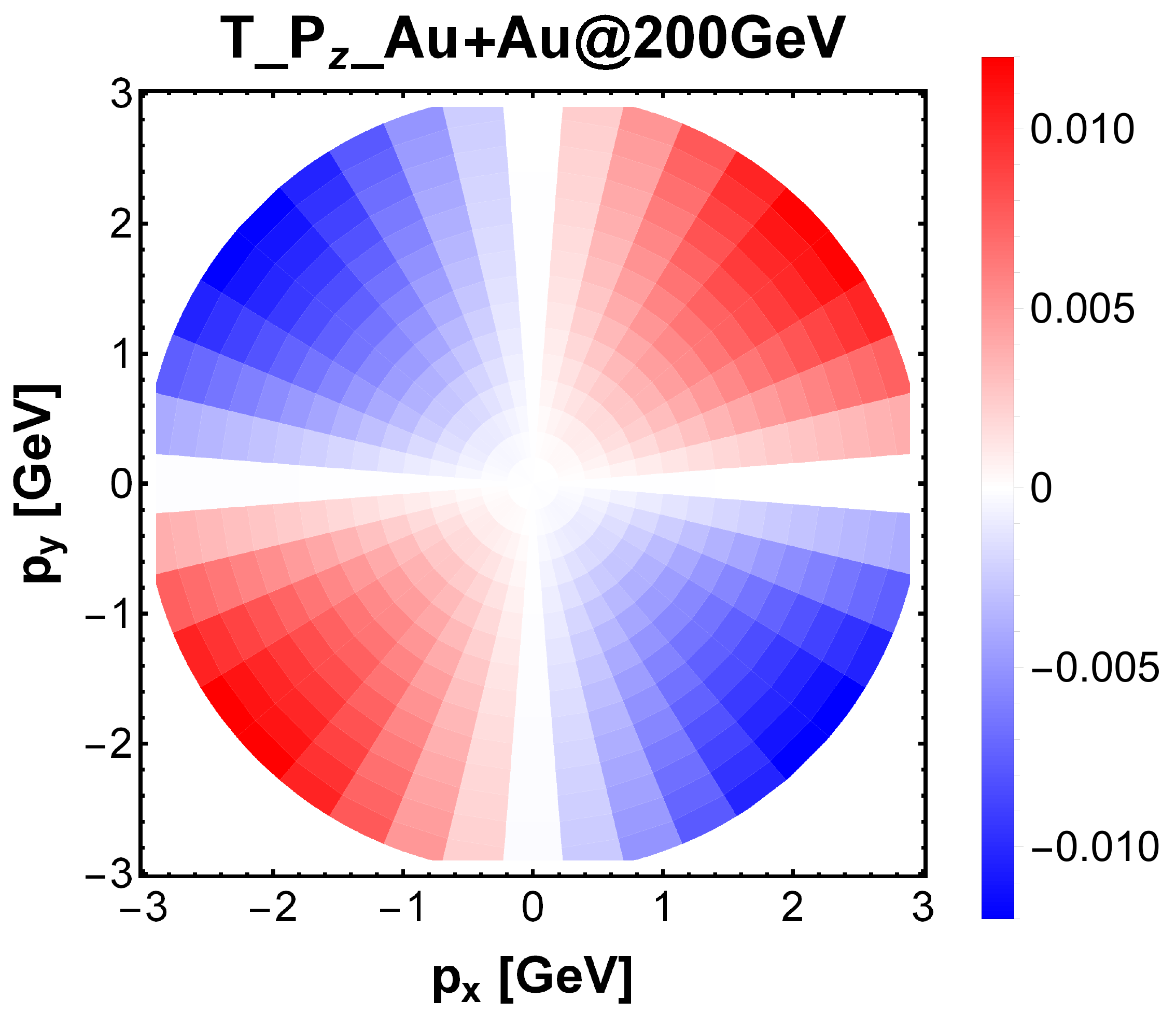}
\includegraphics[scale=0.3]{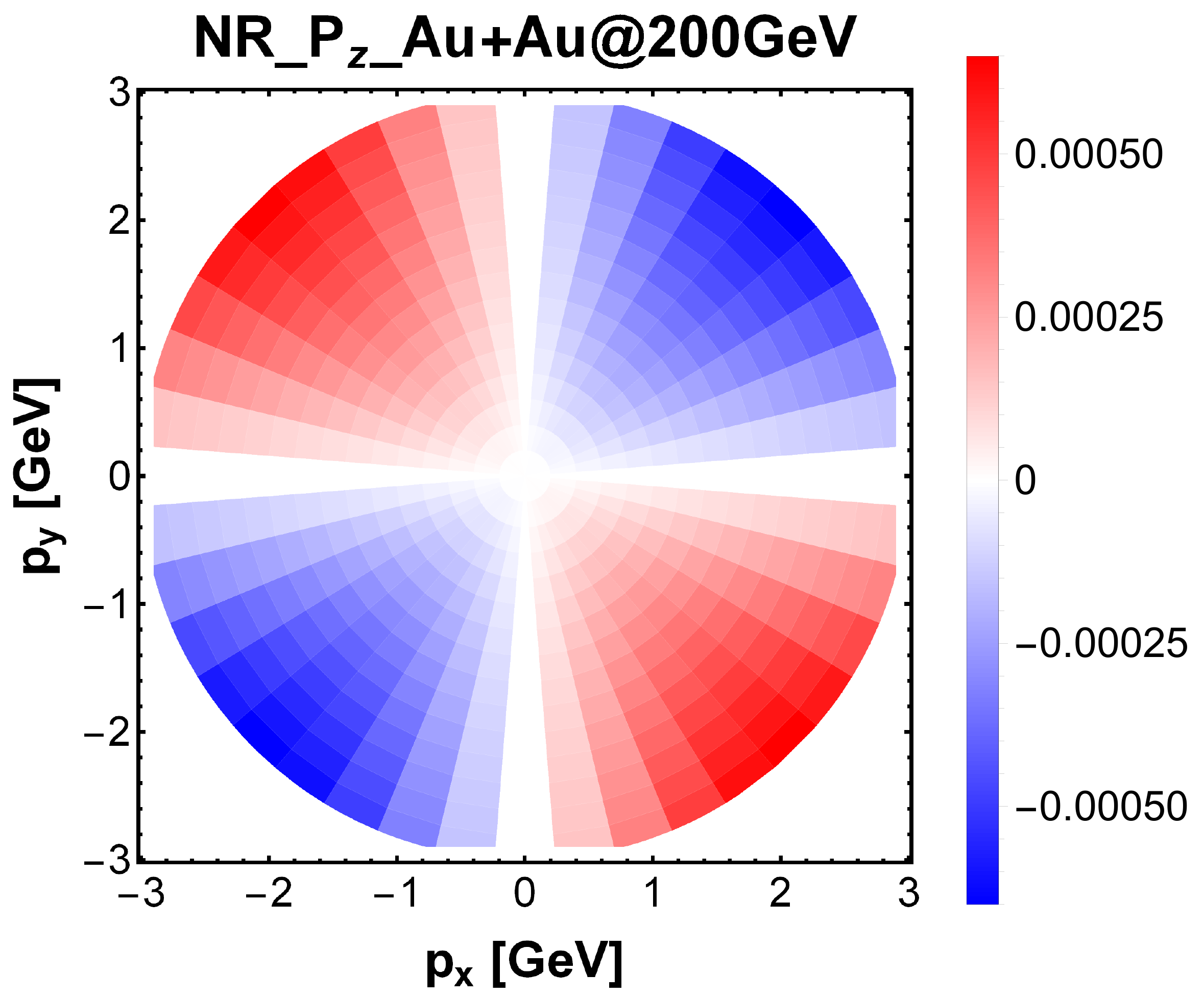}
\caption{The longitudinal polarization in Au+Au collisions at 200 GeV and $Y\in[-1,1]$
with the optical Glauber initial condition as functions of $(p_{x},p_{y})$.
The upper-left, upper-right, lower-left and lower-right panel correspond to the thermal, kinematic, 
T- and NR vorticity, respectively. 
\label{fig:mid-rap-au-au}}
\end{figure}

With the optical Glauber initial condition, we present the results
for the longitudinal polarization for $\Lambda$. Figure \ref{fig:mid-rap-au-au}
shows $\mathcal{P}_{z}^{(i)}(p)$ for four types of spin chemical
potentials or vorticities $i=K,T,\mathrm{th},\mathrm{NR}$. The transverse
momentum $p_{x}$ and $p_{y}$ all range from $-3$ to $3$ GeV. We
see that the T-vorticity has the sign $(+,-,+,-)$ from the first
to fourth quadrants consistent with the data. The kinematic, thermal
and NR vorticity have the sign $(-,+,-,+)$ opposite in comparison
with the data. Note that there is no contribution from the space-time
gradient of the temperature in the kinematic vorticity, the temperature
gradient part in the T-vorticity has the opposite sign to that in the
thermal vorticity. In Fig. \ref{fig:mid-rap-au-au} we see that the
magnitude of $\mathcal{P}_{z}^{(K)}(p)$ is smaller than that of $\mathcal{P}_{z}^{(\mathrm{th})}(p)$
and $\mathcal{P}_{z}^{(T)}(p)$, indicating the dominance of the temperature
gradient parts in the thermal vorticity and T-vorticity.

\begin{figure}[H]
\begin{centering}
\includegraphics[scale=0.3]{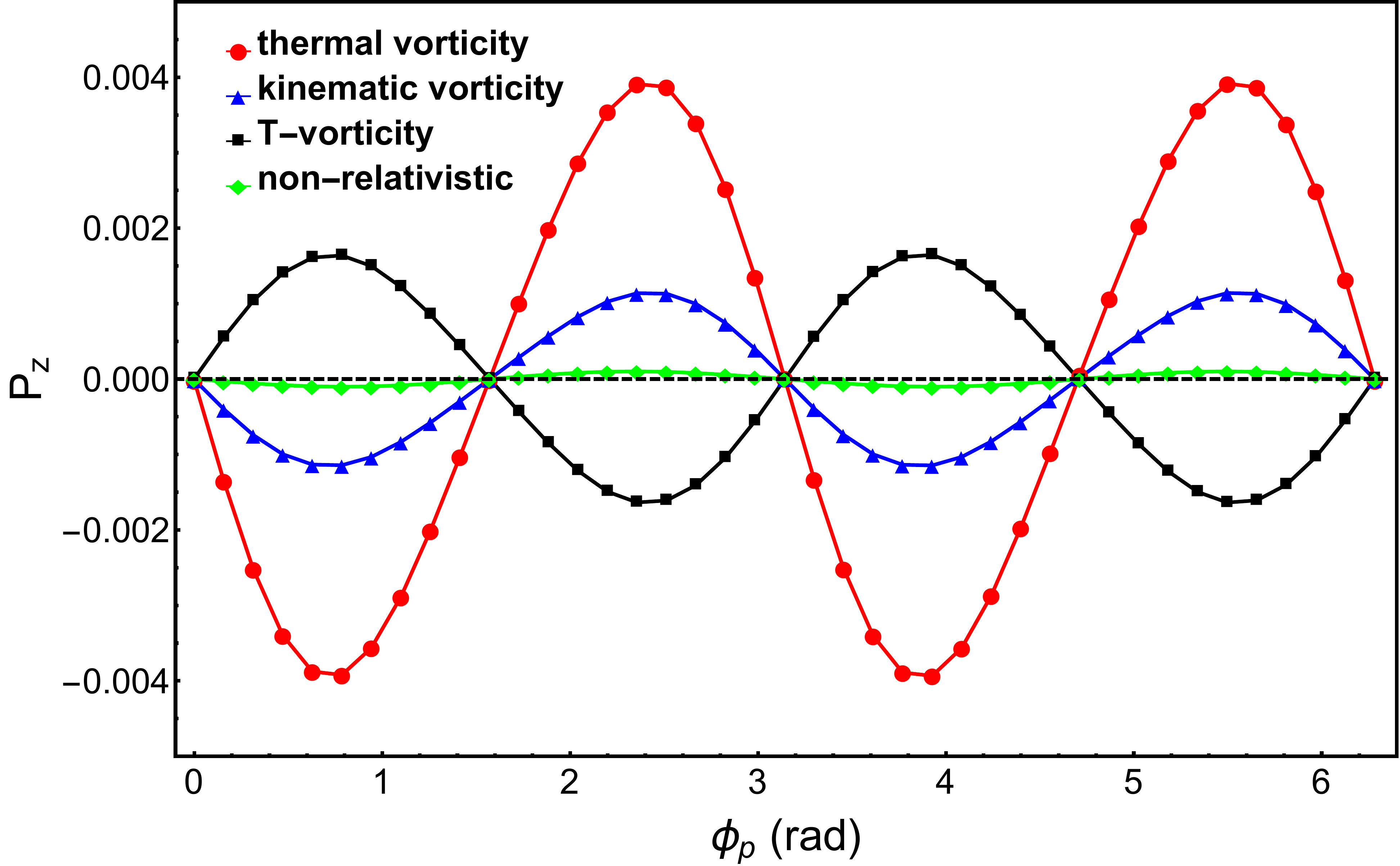}
\par\end{centering}

\begin{centering}
\includegraphics[scale=0.3]{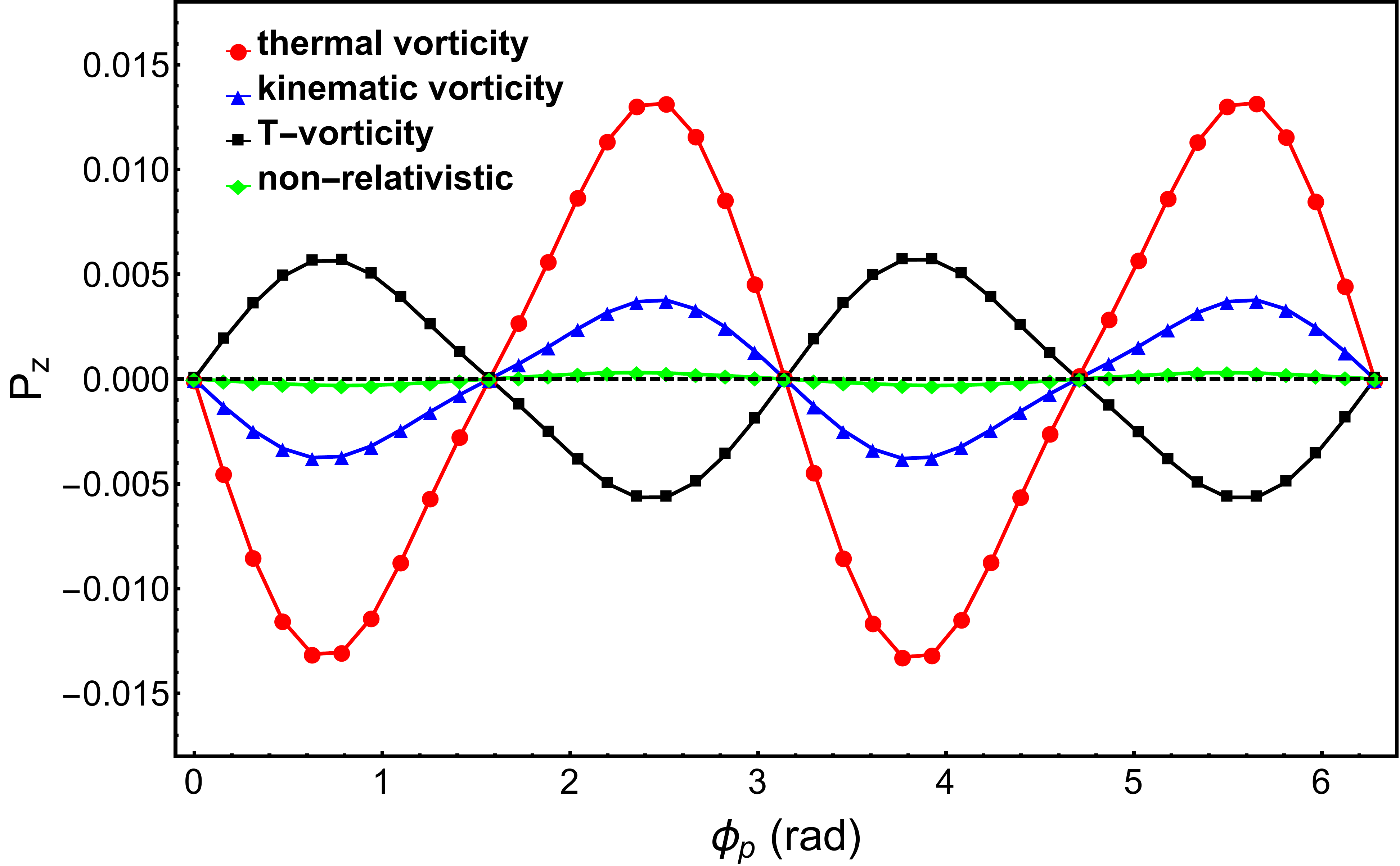}
\par\end{centering}

\caption{The longitudinal polarization as functions of azimuthal angles in
transverse momentum in Au+Au collisions with the optical Glauber initial
condition. Upper panel: $p_{T}\in[0,1.2]$
GeV, lower panel: $[0,3]$ GeV. 
\label{fig:azimuthal-angle-long-polar}}
\end{figure}

The longitudinal polarizations from four types of vorticities as functions
of azimuthal angles in transverse momenta are shown in Fig. \ref{fig:azimuthal-angle-long-polar}.
The azimuthal angle relative to the reaction plane (zx-plane) is defined
as $\tan\phi_{p}=p_{y}/p_{x}$. The azimuthal angle distribution of
the polarization $\overrightarrow{\mathcal{P}}(\phi_{p})$ is obtained
by taking an average over $p_{T}=\sqrt{p_{x}^{2}+p_{y}^{2}}$ for
$\overrightarrow{\mathcal{P}}(p)$
\begin{equation}
\overrightarrow{\mathcal{P}}(\phi_{p})=\frac{1}{\Delta p_{T}}\int_{p_{T}^{\mathrm{min}}}^{p_{T}^{\mathrm{max}}}dp_{T}\overrightarrow{\mathcal{P}}(p),\label{eq:pt-int}
\end{equation}
where $\Delta p_{T}=p_{T}^{\mathrm{max}}-p_{T}^{\mathrm{min}}$ denotes
the range of the transverse momentum. In Fig. \ref{fig:azimuthal-angle-long-polar}
we see that $\mathcal{P}_{z}^{(T)}(\phi_{p})\sim\sin(2\phi_{p})$
which is consistent with the data, while all $\mathcal{P}_{z}^{(i)}(\phi_{p})\sim-\sin(2\phi_{p})$
with $i=K,\mathrm{th},\mathrm{NR}$ which have the wrong sign in comparison
with the data. The magnitude of $\mathcal{P}_{z}^{(\mathrm{th})}(\phi_{p})$
is the largest since it is the sum of the kinematic vorticity contribution
and the temperature gradient contribution which have the same sign.
But in $\mathcal{P}_{z}^{(T)}(\phi_{p})$ the kinematic vorticity
and temperature gradient contribution have the opposite sign and the
latter is dominant over the former. This is the reason that $\omega_{\mu\nu}^{(T)}(T)$
and $\omega_{\mu\nu}^{(\mathrm{th})}(T)$ have the opposite sign as
shown in Eqs. (\ref{T vorticity},\ref{thermal vorticity}).

Since there is no initial OAM in the optical Glauber initial condition,
the polarizations in the $y$ direction $\mathcal{P}_{y}(p)$ are
vanishing for all four types of vorticities.

\subsection{Results with AMPT initial condition}

In this subsection, we present the results for the AMPT initial condition
which encodes the initial OAM of two nuclei in $-y$ direction.

\begin{figure}[H]
\begin{centering}
\includegraphics[scale=0.3]{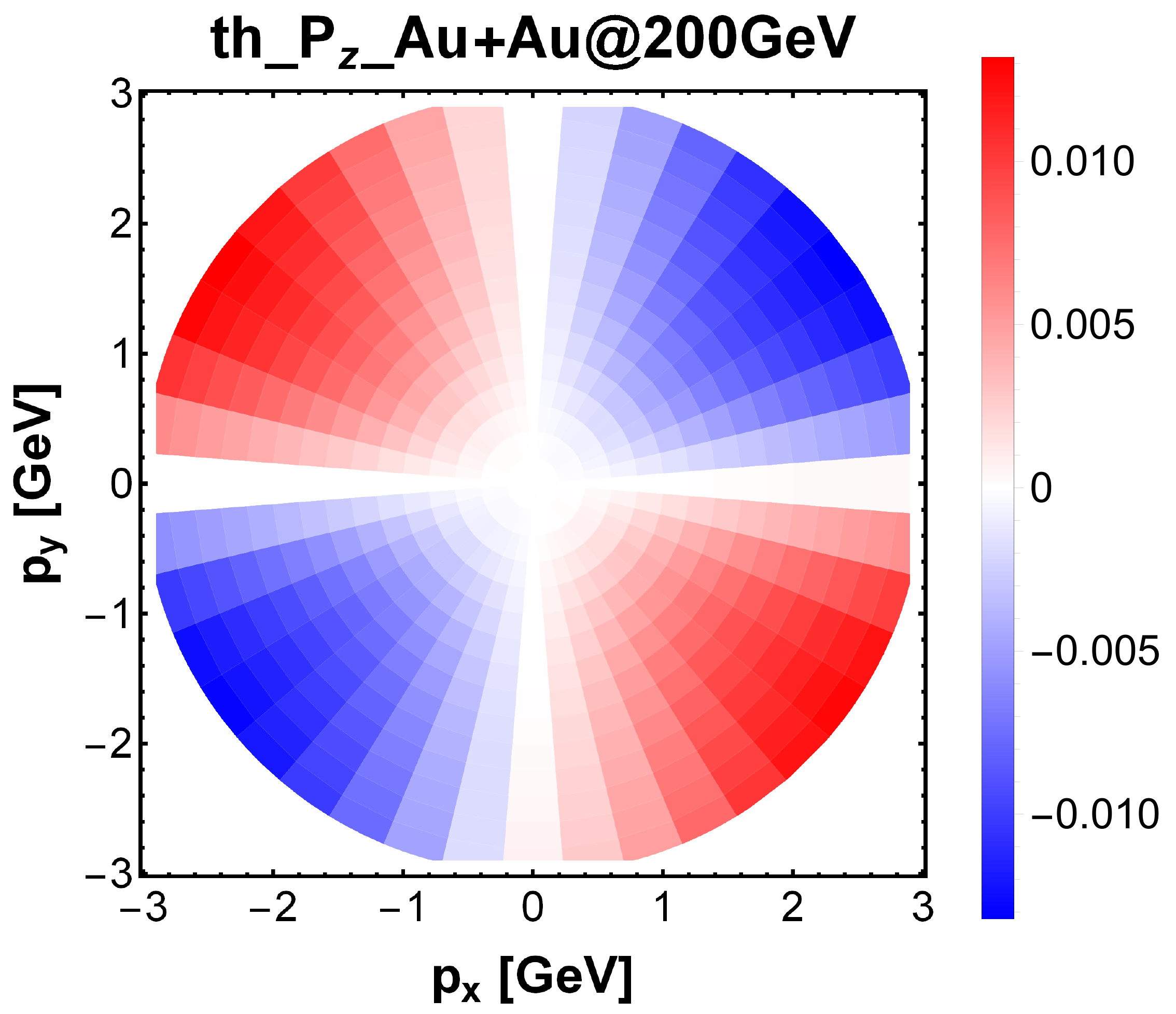}\includegraphics[scale=0.3]{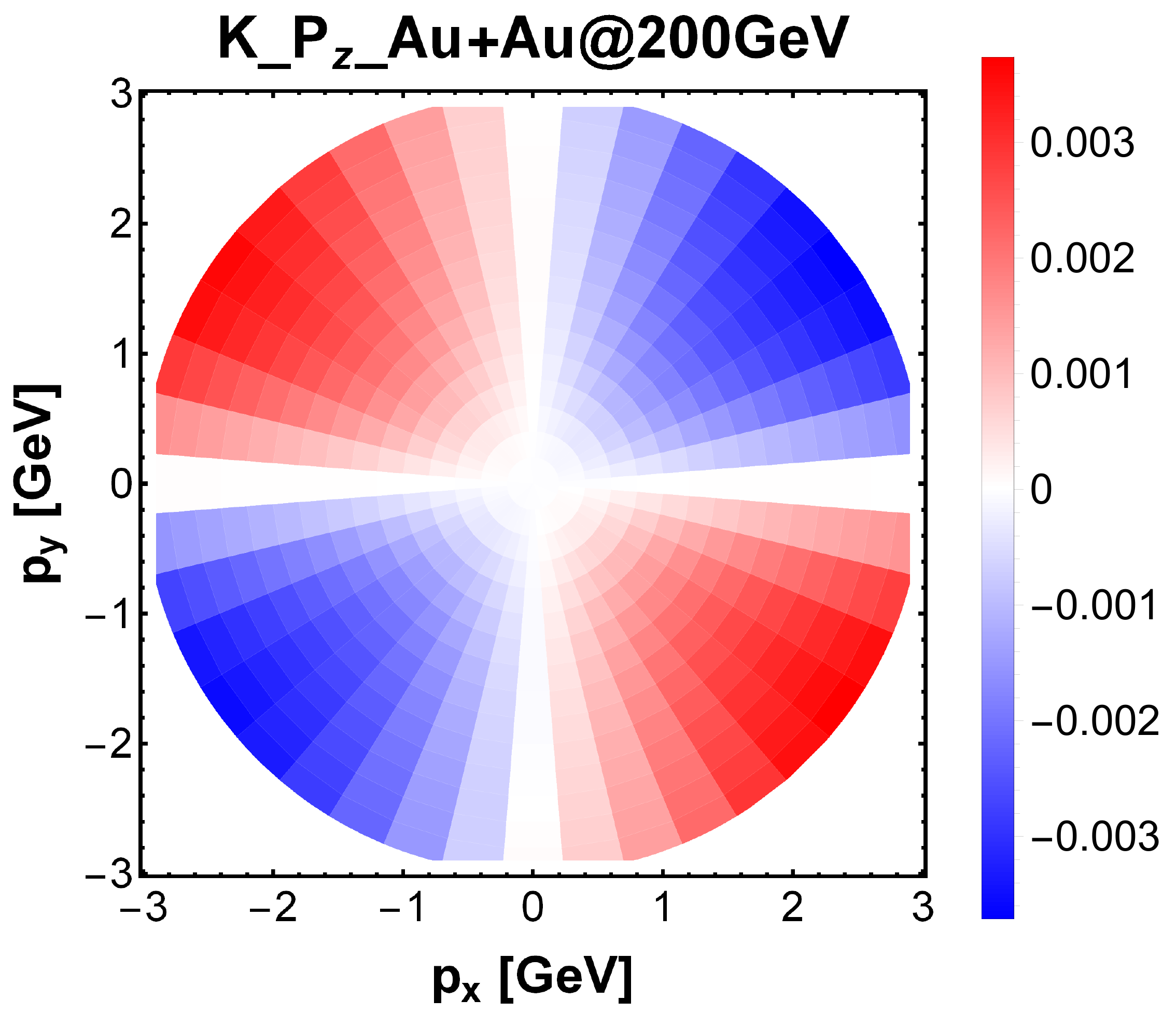}
\par\end{centering}

\begin{centering}
\includegraphics[scale=0.3]{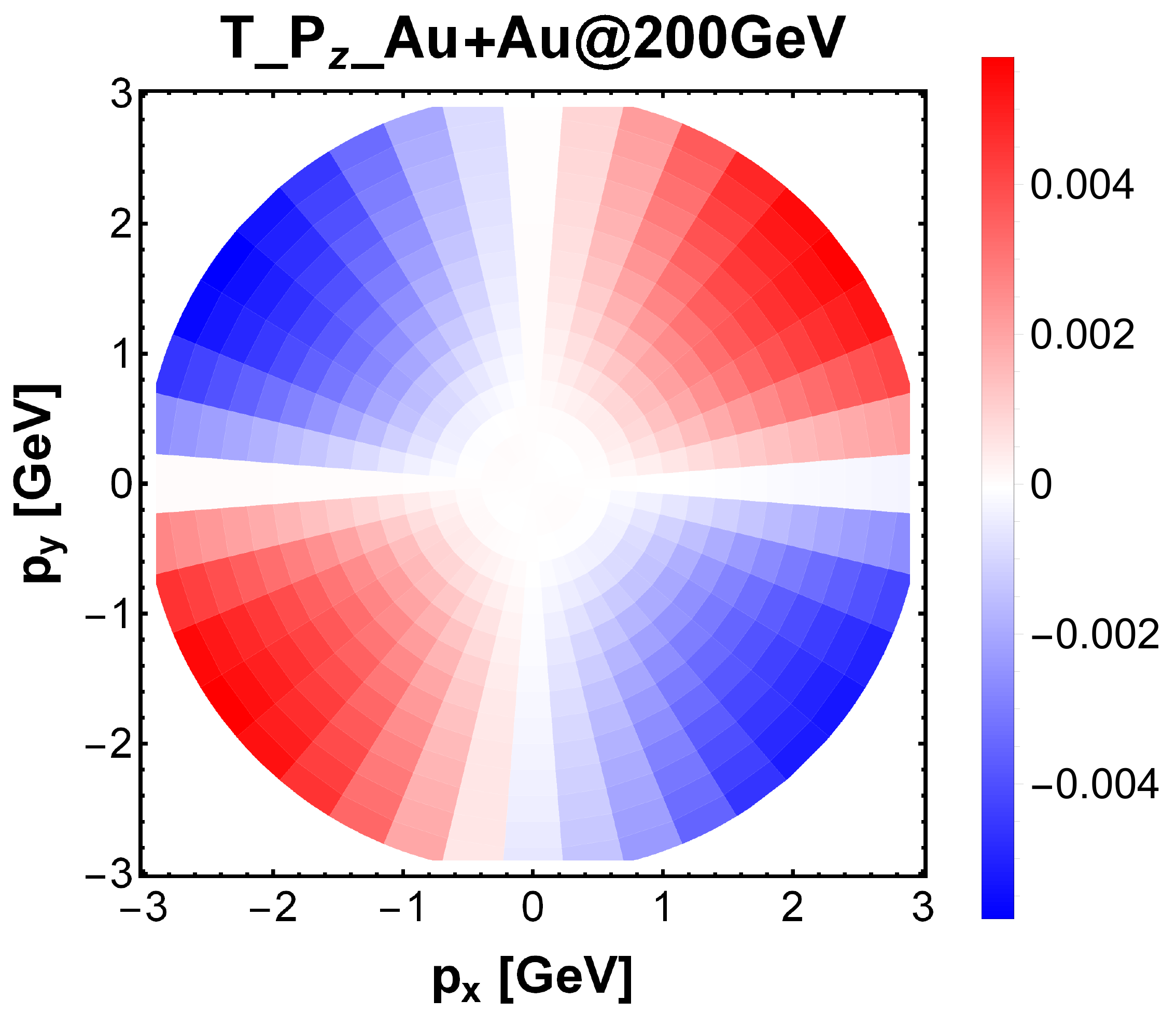}\includegraphics[scale=0.3]{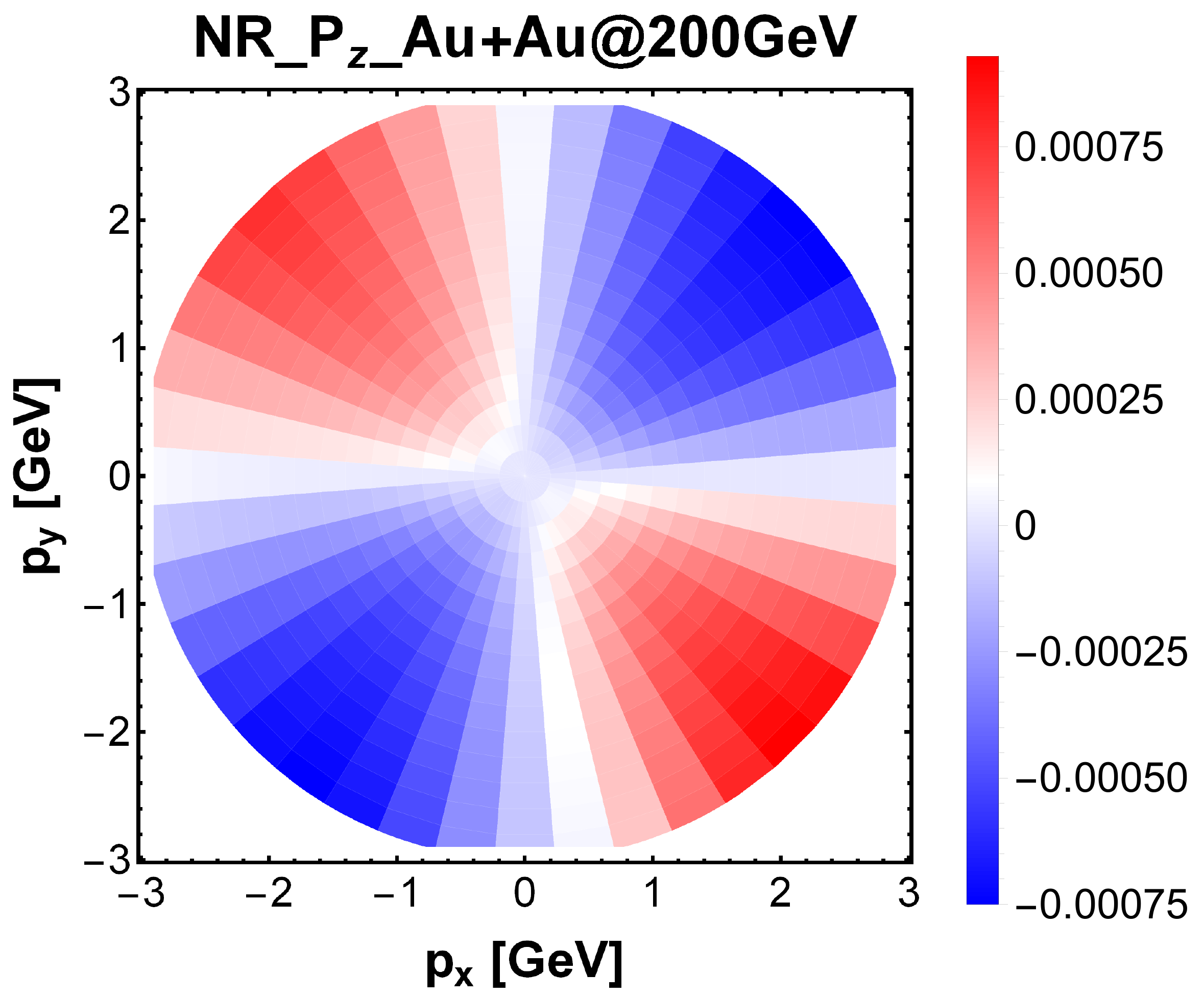}
\par\end{centering}

\caption{The longitudinal polarization in Au+Au collisions at 200 GeV in the
rapidity range $Y\in[-1,1]$ with the AMPT initial condition as functions
of $(p_{x},p_{y})$. 
The upper-left, upper-right, lower-left and lower-right panel correspond to the thermal, kinematic, 
T- and NR vorticity, respectively.
\label{fig:pz-ampt}}
\end{figure}

\begin{figure}[H]
\begin{centering}
\includegraphics[scale=0.3]{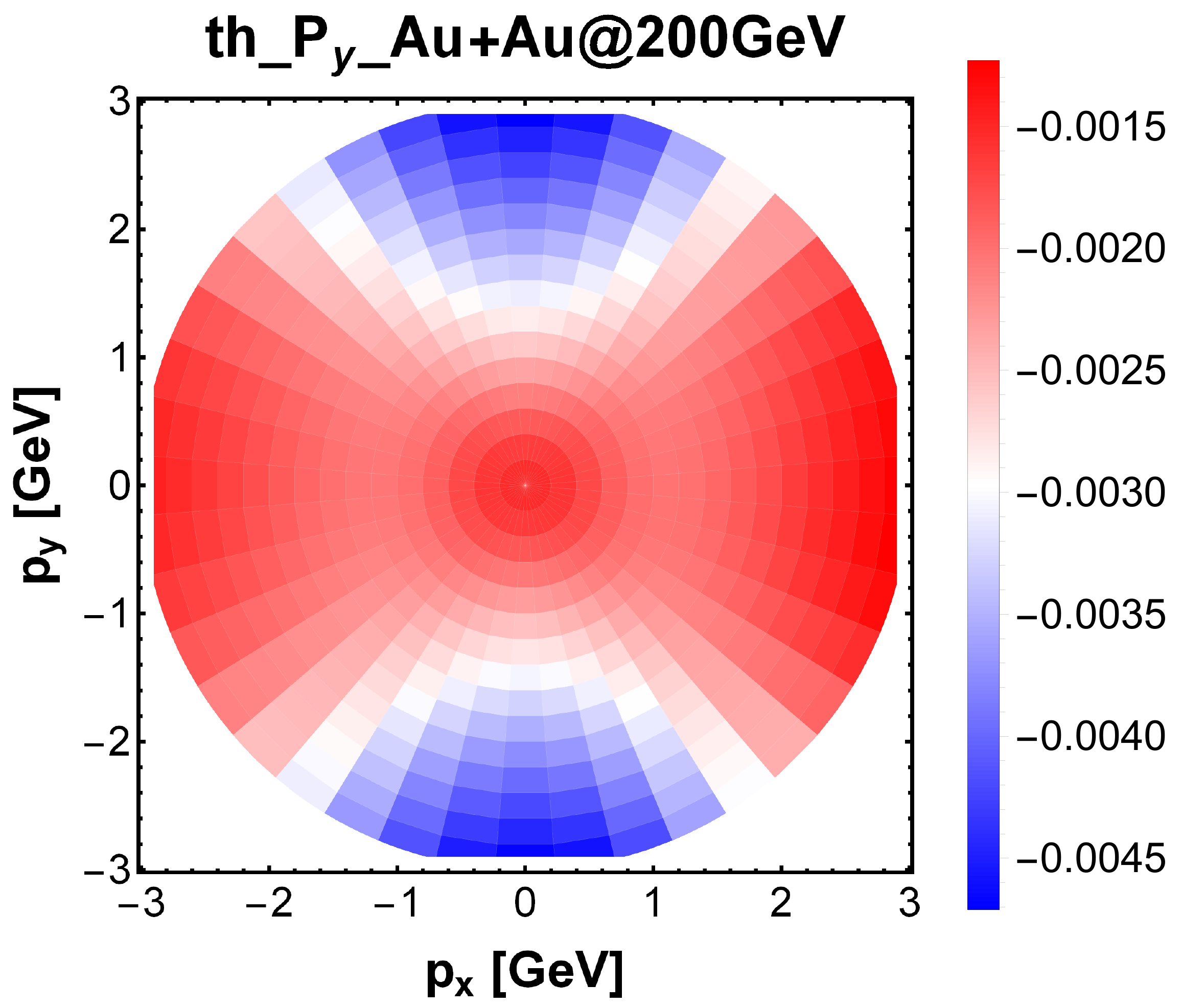}\includegraphics[scale=0.3]{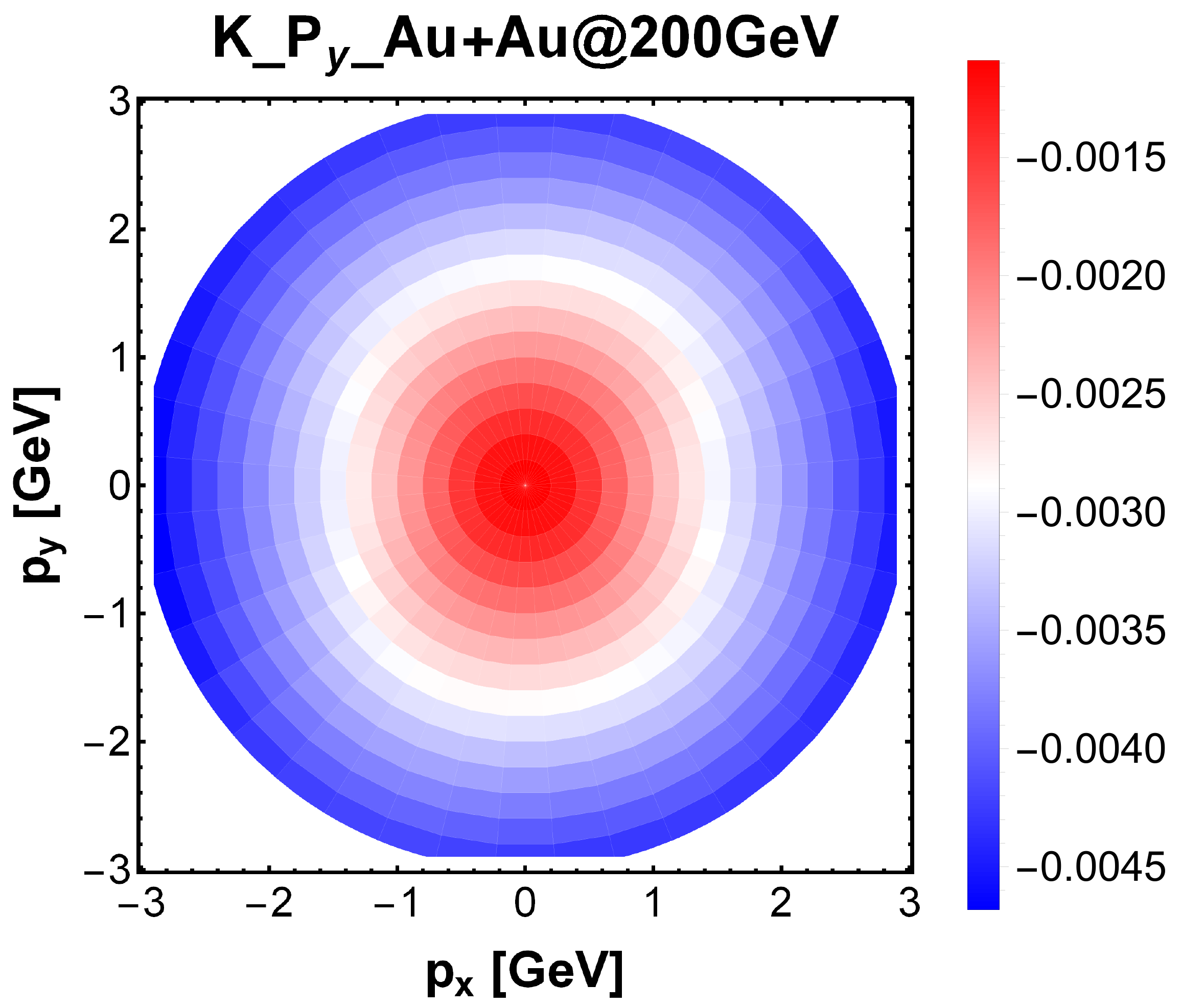}
\par\end{centering}

\centering{}
\includegraphics[scale=0.3]{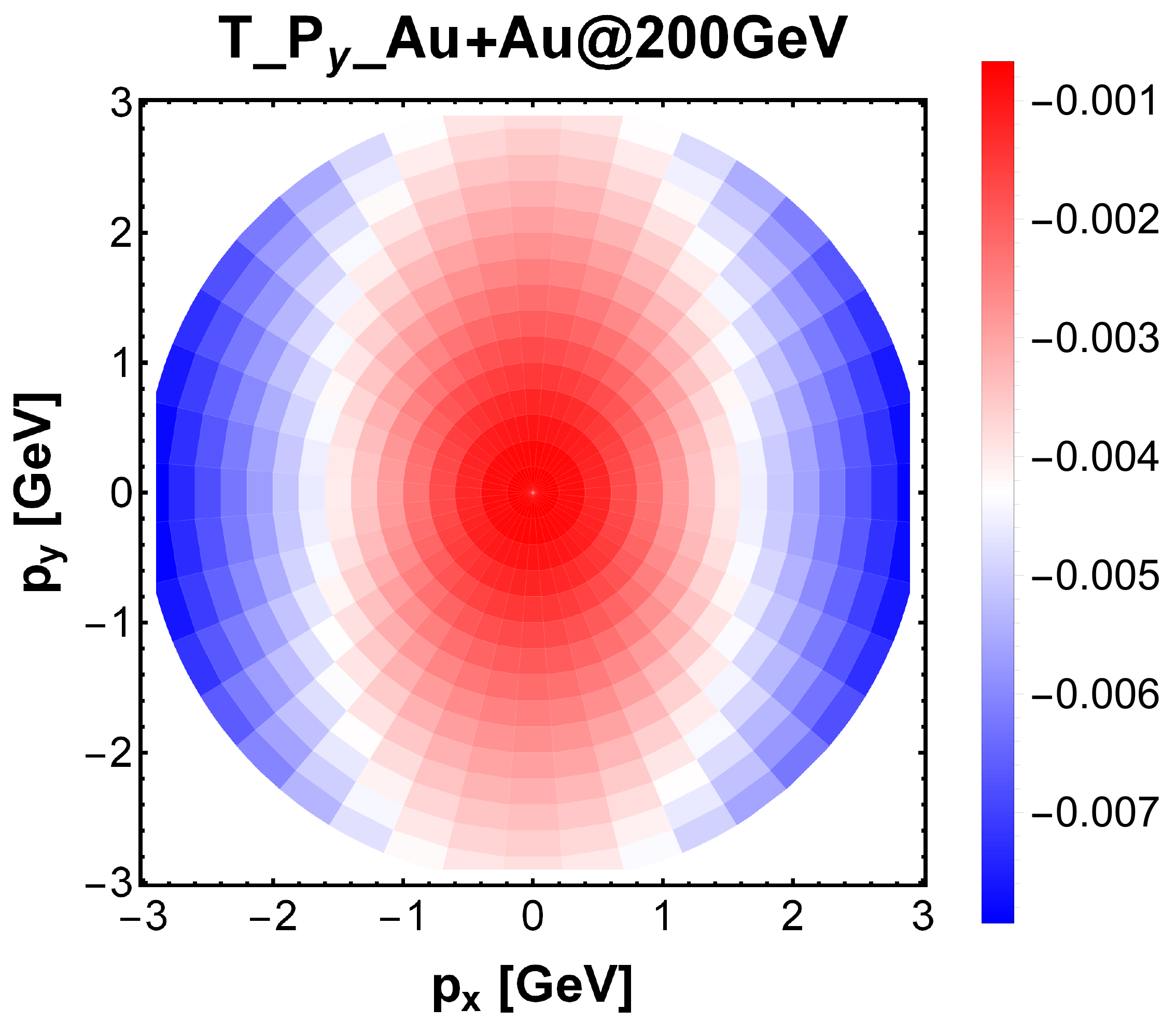}\includegraphics[scale=0.3]{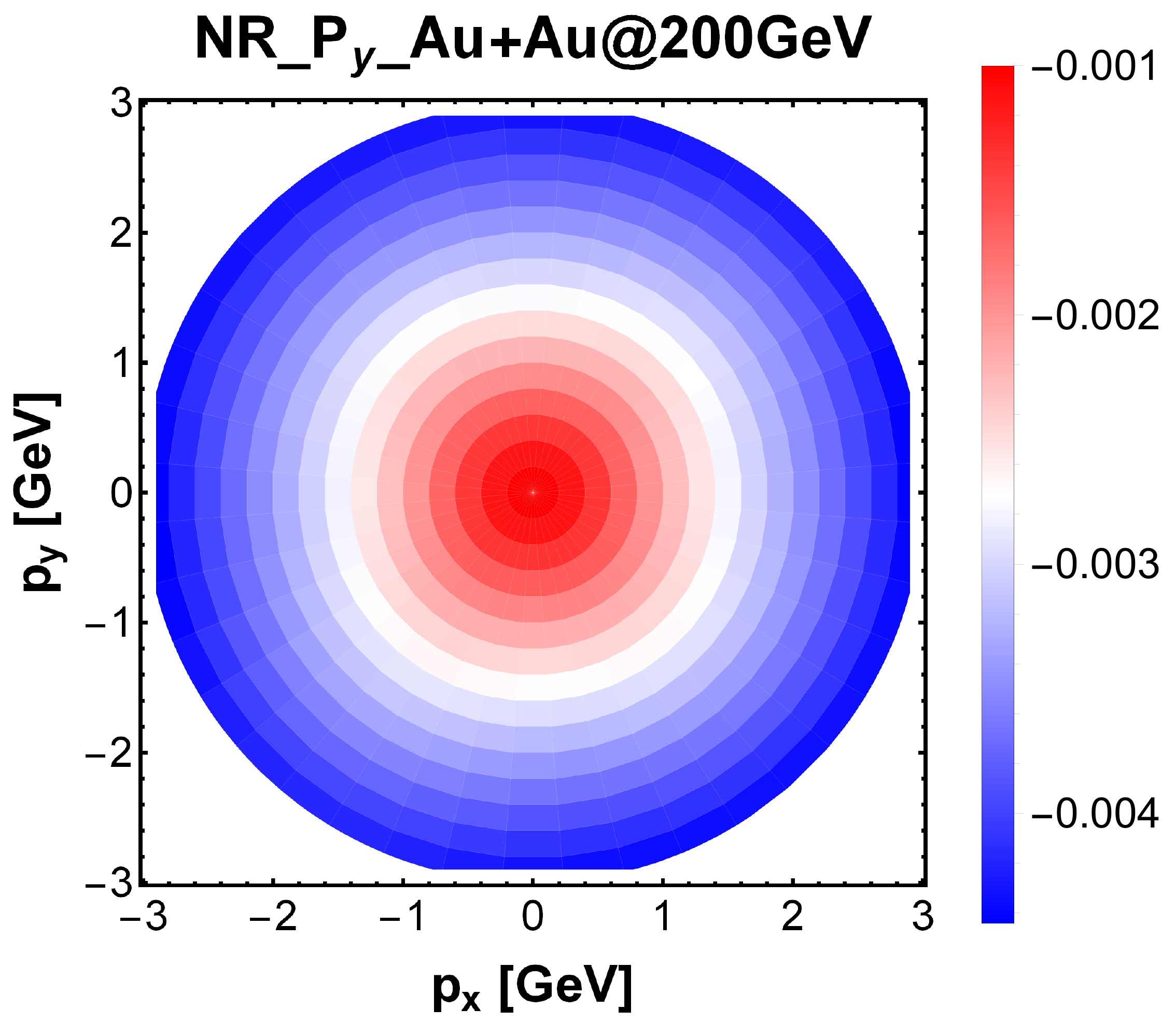}
\caption{The polarization in $-y$ direction. All other kinematic conditions
are the same as in Fig. \ref{fig:pz-ampt}. \label{fig:py-ampt}}
\end{figure}

The results of $\mathcal{P}_{z}(p)$ and $\mathcal{P}_{y}(p)$ for
four types of vorticities are shown in Fig. \ref{fig:pz-ampt} and
Fig. \ref{fig:py-ampt} respectively. We see in Fig. \ref{fig:pz-ampt}
that the signs of $\mathcal{P}_{z}(p)$ with the AMPT initial condition
are the same as those with the Glauber initial condition but the magnitudes
of $\mathcal{P}_{z}(p)$ with the AMPT initial condition are smaller
than those with the Glauber initial condition except $\mathcal{P}_{z}(p)$ 
for the NR vorticity that has almost the same magnitude with both initial conditions. 

\begin{figure}[H]
\begin{centering}
\includegraphics[scale=0.3]{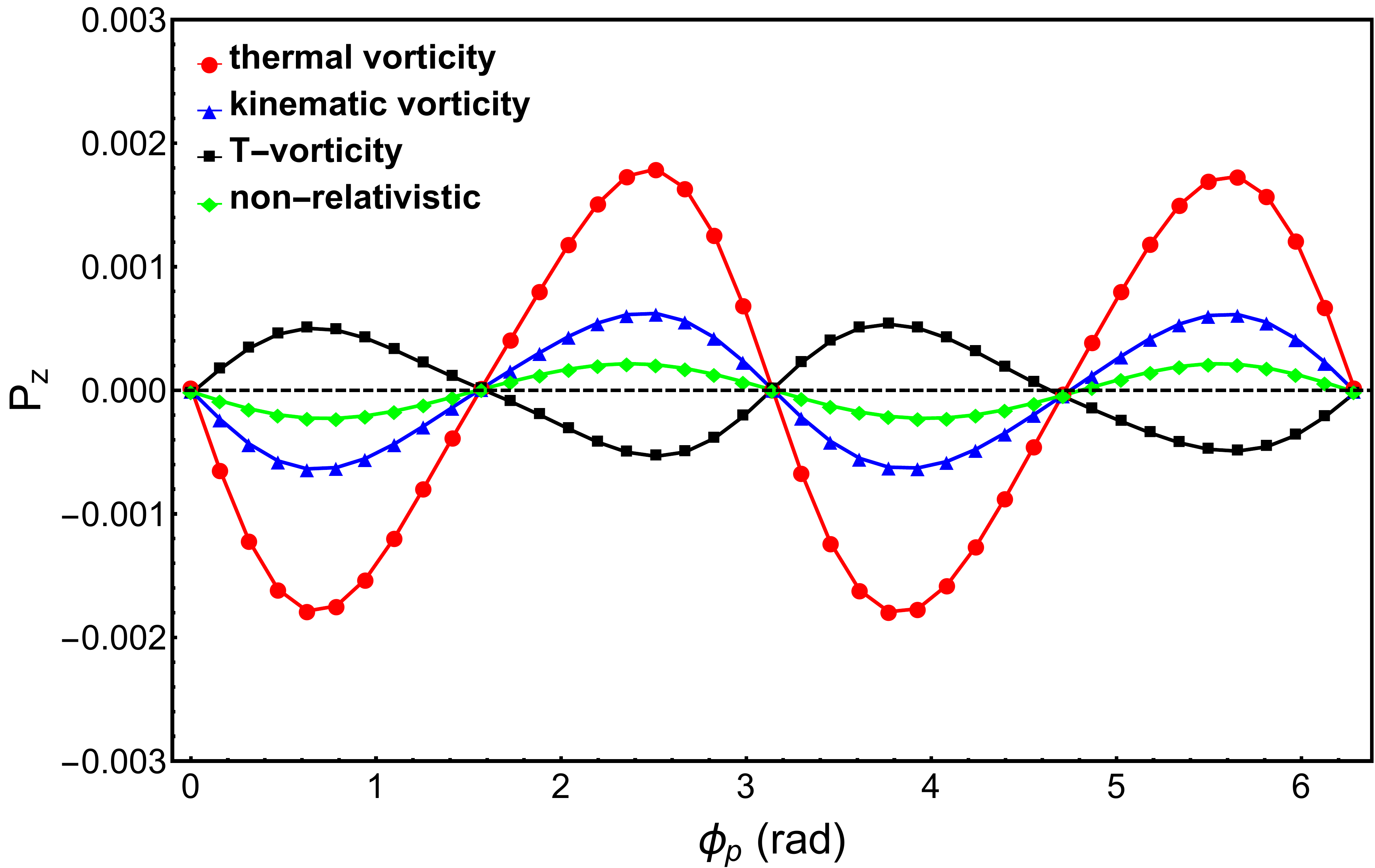}
\par\end{centering}

\centering{}\includegraphics[scale=0.3]{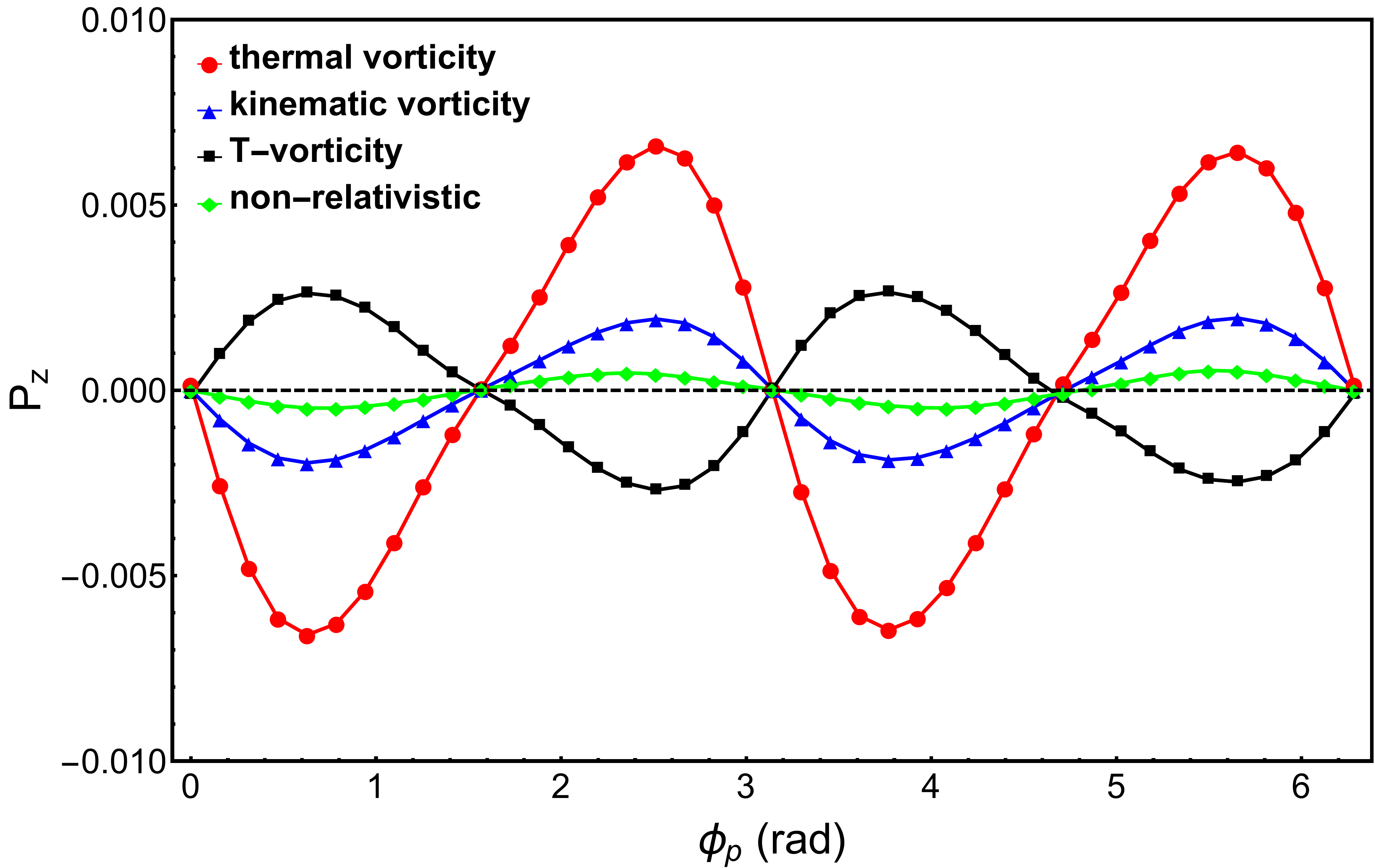}
\caption{The longitudinal polarization as functions of azimuthal angles in
transverse momentum in Au+Au collisions with the AMPT initial condition.
Upper panel: $p_{T}\in[0,1.2]$ GeV, lower panel: $p_{T}\in[0,3]$
GeV. \label{fig:pz-ampt-phi}}
\end{figure}

We can take an average over $p_{T}$ for $\mathcal{P}_{z}(p)$ in
a transverse momentum range to obtain $\mathcal{P}_{z}(\phi_{p})$.
The results for $\mathcal{P}_{z}(\phi_{p})$ are shown in Fig. \ref{fig:pz-ampt-phi}
for two transverse momentum ranges. We see that the magnitudes of
$\mathcal{P}_{z}(\phi_{p})$ become smaller in the $p_{T}$ range
with smaller transverse momenta. For the range $p_{T}\in[0,1.2]$
GeV, the magnitude of $\mathcal{P}_{z}(\phi_{p})$ matches the data.
But if we choose $p_{T}\in[0,3]$ GeV, the magnitude of $\mathcal{P}_{z}(\phi_{p})$
is one order of magnitude larger than the data.

\begin{figure}[H]
\centering{}\includegraphics[scale=0.3]{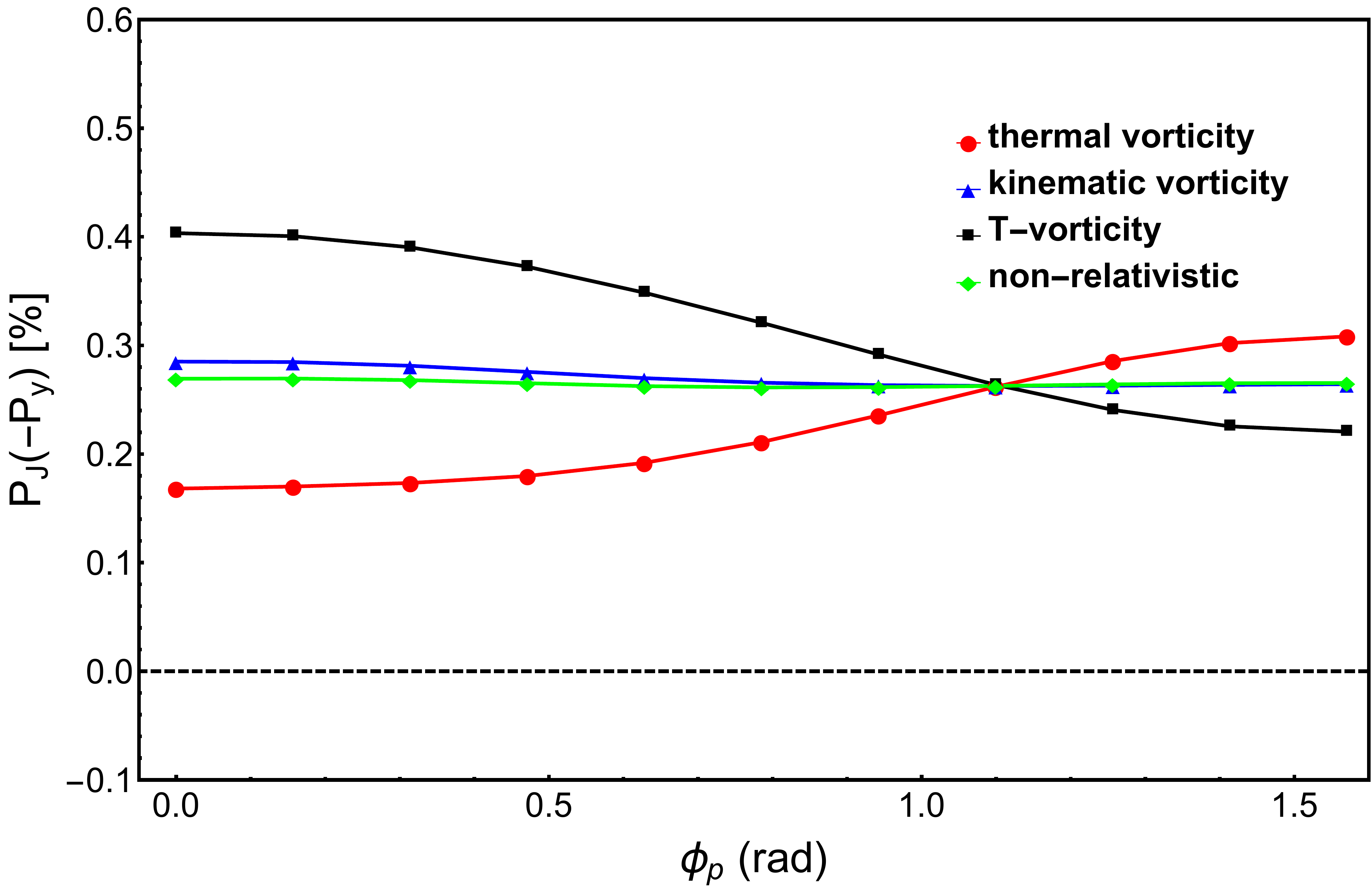}
\caption{The polarization in $-y$ direction as functions of azimuthal angles
in transverse momentum in Au+Au collisions with the AMPT initial condition.
The transverse momentum range is set to $p_{T}\in[0,3]$ GeV. \label{fig:py-ampt-phi}}
\end{figure}

In contrast to the vanishing $\mathcal{P}_{y}(p)$ with the Glauber
initial condition, we obtain finite values of $\mathcal{P}_{y}(p)$
in the AMPT initial condition as shown in Fig. \ref{fig:py-ampt}.
The results for $\mathcal{P}_{y}(\phi_{p})$ are displayed in Fig.
\ref{fig:py-ampt-phi}. All four types of vorticities give the correct
sign of the initial OAM in $-y$ direction. Note that only $\mathcal{P}_{y}^{(T)}(\phi_{p})$
for the T-vorticity gives the falling trend in $\phi_{p}$ consistent
with the data. Although we have the correct trend in $\phi_{p}$
in $\mathcal{P}_{y}^{(T)}(\phi_{p})$, our results fall slower 
than the data as $\phi_{p}$ increases. Our results for $\mathcal{P}_{y}^{(T)}(\phi_{p})$
match the data at $\phi_{p}=0$, but at $\phi_{p}=\pi/2$ our results
are $\mathcal{P}_{y}^{(T)}(\phi_{p})\approx0.25$ while the data approach
zero.

\begin{figure}[H]
\centering{}\includegraphics[scale=0.3]{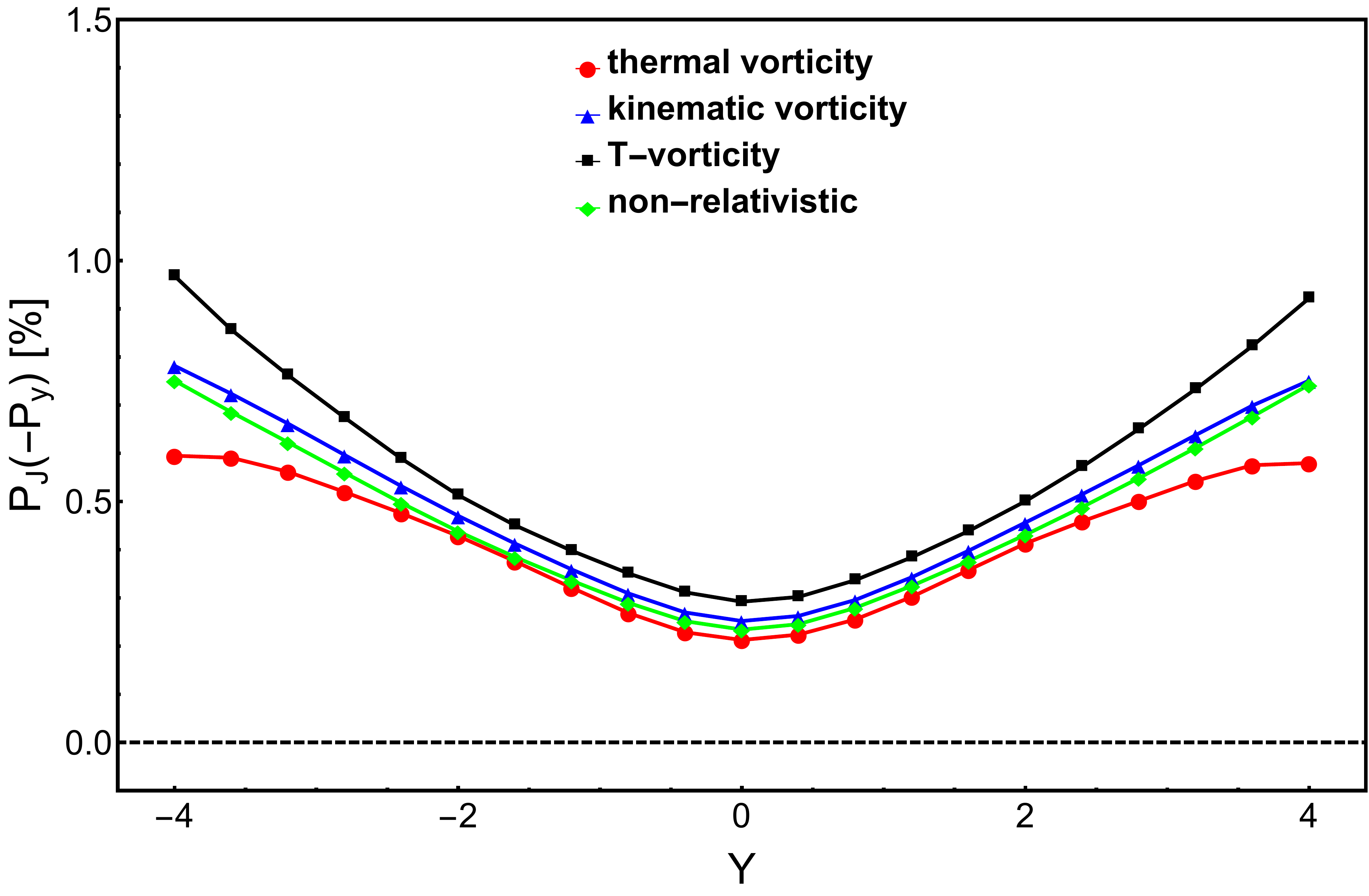}
\caption{The polarization in $-y$ direction as functions of the rapidity in
Au+Au collisions with the AMPT initial condition. The transverse momentum
range is set to $p_{T}\in[0,3]$ GeV. \label{fig:py-rapidity}}
\end{figure}

We also calculated $\mathcal{P}_{y}(Y)$ in $-y$ direction as functions
of the rapidity $Y$, the results are shown in Fig. \ref{fig:py-rapidity}.
We see that $\mathcal{P}_{y}(Y)$ is an even function of $Y$ and increases slowly with $|Y|$.
The values of $\mathcal{P}_{y}(Y)$ are very close for the kinematic
and NR-vorticity, and $\mathcal{P}_{y}(Y)$ for the T-vorticity is the largest while that for thermal vorticity is the smallest.

\subsection{With different average method in momentum}

\begin{figure}[H]
\begin{centering}
\includegraphics[scale=0.3]{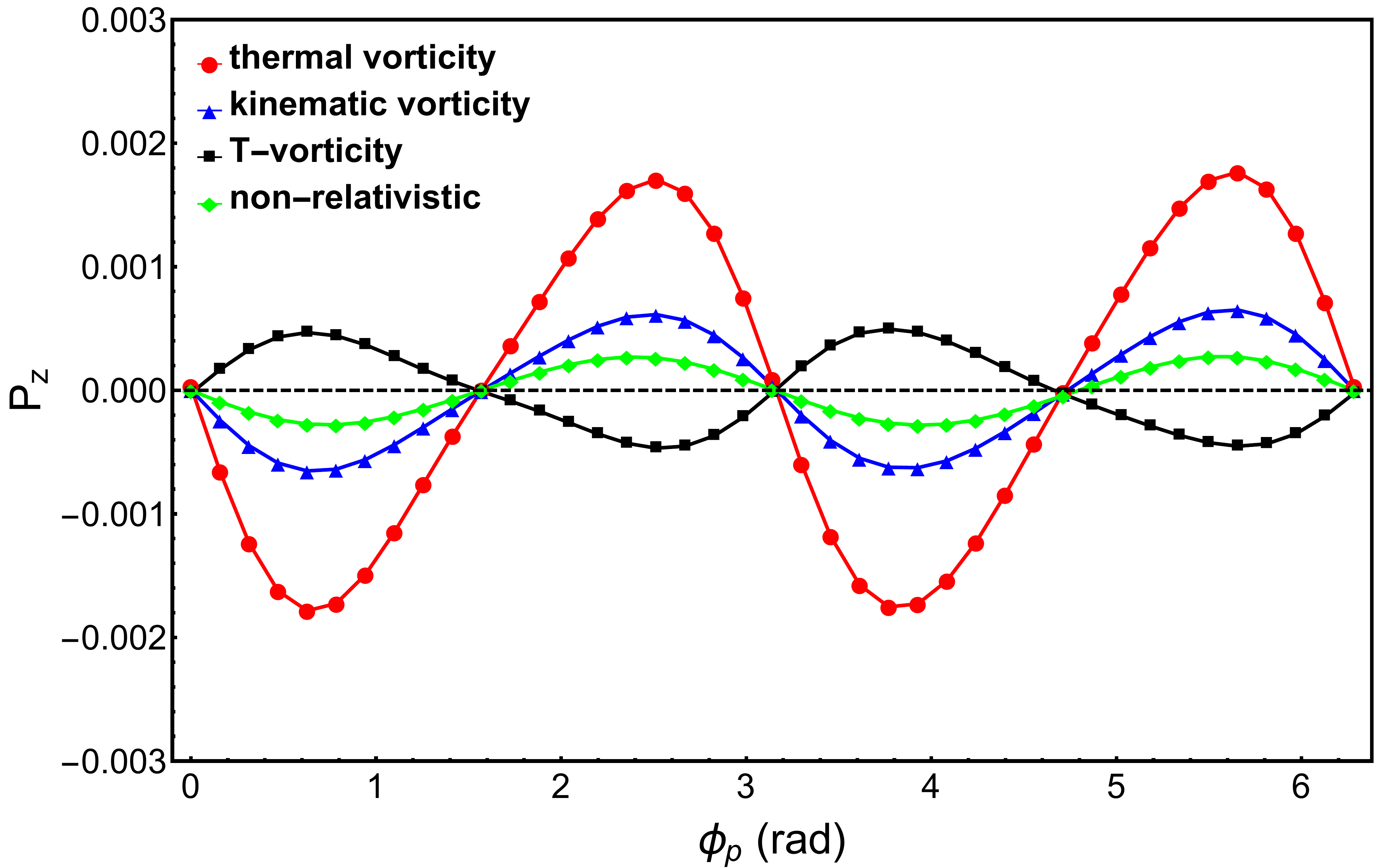}
\par\end{centering}

\caption{The longitudinal polarization as functions of azimuthal angles in
transverse momentum in Au+Au collisions with the AMPT initial condition.
An alternative average method corresponding to Eq. (\ref{average-spin-vector-1})
is used. The $p_{T}$ range is chosen to be $p_{T}\in[0,1.2]$ GeV
to match the magnitude of the data. \label{fig:pz-ampt-phi-1}}
\end{figure}

We can also choose a different method of the average over transverse
momenta and rapidity to replace Eqs. (\ref{eq:rapidity-av},\ref{eq:pt-int}).
From Eq. (\ref{average-spin-vector}) we can take an average of the
denominator and numerator separately to obtain the $i$-th component
of the polarization vector,
\begin{equation}
\mathcal{P}_{i}(\phi_{p})=-\frac{1}{4m}\epsilon^{i\rho\sigma\tau}\frac{\int_{p_{T}^{\mathrm{min}}}^{p_{T}^{\mathrm{max}}}dp_{T}p_{T}\int_{-\Delta Y/2}^{\Delta Y/2}dY\int d\Sigma_{\lambda}p^{\lambda}p_{\tau}\Omega_{\rho\sigma}f_{FD}(1-f_{FD})}{\int_{p_{T}^{\mathrm{min}}}^{p_{T}^{\mathrm{max}}}dp_{T}p_{T}\int_{-\Delta Y/2}^{\Delta Y/2}dY\int d\Sigma_{\lambda}p^{\lambda}f_{FD}}+O(\Omega_{\mu\nu}^{2}).\label{average-spin-vector-1}
\end{equation}
Note that we have introduced an additional $p_{T}$ factor into the
$p_{T}$ integrals in both the denominator and numerator since it
corresponds to the Lorentz invariant integral $d^{3}p/E_{p}$. The
numerical results for $\mathcal{P}_{z}(\phi_{p})$ are presented in
Fig. \ref{fig:pz-ampt-phi-1}. We see that with the same cutoffs for
$p_{T}$, the results for $\mathcal{P}_{z}(\phi_{p})$ from Eq. (\ref{average-spin-vector-1})
are a little larger than from Eqs. (\ref{eq:rapidity-av},\ref{eq:pt-int}).
The same behavior also occurs in the results for $\mathcal{P}_{y}(\phi_{p})$
with two different average methods.

\section{Discussions}

We make some remarks about the results. We have checked the first
two conditions in (\ref{eq:conditions}) and found that they are not
fulfilled, so the use of the thermal vorticity as the spin chemical
potential is not justified in the hydro-simulation.

For $\mathcal{P}_{y}(\phi_{p})$, we find that only the T-vorticity gives the right trend in $\phi_{p}$ comparing to the data, although it decreases slower than the data. Except the trend in $\phi_{p}$, all vorticities can give the global polarization consistent with the data. The reason why the T-vorticity can give the right trend in $\phi_{p}$ may be understood as follows. The T-vorticity is conserved so that the T-vorticity flux is frozen in the fluid and move with the fluid cell. In this sense, we can regard the T-vorticity flux as a kind of conserved charge. At the early stage of a non-central collision, the T-vorticity in the out-of-plane direction may be induced by the global OAM, then as the pressure gradient is stronger in the in-plane direction than the out-of-plane direction, the T-vorticity will have a positive elliptic flow which results in the unique $\phi_{p}$ dependence as shown in Fig.~\ref{fig:py-ampt-phi}. This suggests that if the spin is (quasi-)conserved, after polarized in the early stage by the OAM, the pressure gradient would lead to a similar $\phi_{p}$ dependence as that for the T-vorticity. This may be verified by the simulation using spin hydrodynamics~\cite{Florkowski:2017ruc,Hattori:2019lfp}.

We see very different and even opposite behaviors of $\mathcal{P}_{z}(\phi_{p})$
from different vorticities. This might be related to the fact that
$\mathcal{P}_{z}(\phi_{p})$ is one order magnitude smaller than $\mathcal{P}_{y}(\phi_{p})$
since there is no initial OAM in the $z$ direction. Also $\mathcal{P}_{z}(\phi_{p})$
is almost independent of $\mathcal{P}_{y}(\phi_{p})$. This can be
seen from the observation that the results of $\mathcal{P}_{z}(\phi_{p})$
from all types of vorticities in the Glauber initial condition (without
initial OAM) have the same behaviors as in the AMPT initial condition
(with initial OAM). In the optical Glauber initial condition, we found
that $\mathcal{P}_{y}(\phi_{p})$ from all types of vorticities are
vanishing since there is no orbital angular momentum encoded in the
initial state.

Only the T-vorticity in our simulation can describe the data of $\mathcal{P}_{z}(\phi_{p})$
which is the main finding of the paper.
The temperature part $\omega_{\mu\nu}^{(T)}(T)$ in the T-vorticity
(\ref{T vorticity}) plays an essential role in producing the right
sign of $\mathcal{P}_{z}(\phi_{p})$: the sign of $\omega_{\mu\nu}^{(T)}(T)$
is different from $\omega_{\mu\nu}^{(K)}$ but with larger magnitude,
so the T-vorticity takes the sign of $\omega_{\mu\nu}^{(T)}(T)$.
It is just the opposite way for the thermal vorticity (\ref{thermal vorticity})
to make its sign: the temperature part $\omega_{\mu\nu}^{(\mathrm{th})}(T)$
has the same sign as $\omega_{\mu\nu}^{(K)}$.

The implication of the T-vorticity by the data may possibly indicate:
(1) The time behavior of the temperature at the freeze-out is essential
for the T-vorticity to reproduce the correct sign of $\mathcal{P}_{z}(\phi_{p})$.
(2) The T-vorticity might be coupled
with the spin in a similar way that a magnetic moment is coupled
to a magnetic field. Considering an ideal fluid without a conserved charge density 
(such as the baryon  number density)
which is the case in the current hydro-simulation for high energy
heavy ion collisions, $Tu^{\mu}$ can be regarded as a vector potential and 
the T-vorticity tensor is then the corresponding field strength tensor, 
so the conservation of T-vorticity flux is similar to the conservation of 
the magnetic flux in an ideally conducting fluid, see Eqs. (\ref{eq:euler2},\ref{eq:cons-tu}). 
However, such a picture is not yet rigorously established and 
it is also unclear how the roles of T-vorticity and thermal vorticity change 
when the system approaches global equilibrium. Nevertheless, for collisions at lower energies
in which the baryon number density is finite, the conservation of the T-vorticity
flux does not hold anymore \cite{Gao:2014coa}. Thus, the behavior of $\mathcal{P}_{z}(\phi_{p})$
in low energy collisions might provide a test of this point of view.
(3) The assumption that the spin chemical potential can be constructed using $T$ and $u^\mu$ might not be correct, so the fact that the T-vorticity can qualitatively reproduce the experimental data for $\mathcal{P}_{z}(\phi_{p})$ and $\mathcal{P}_{y}(\phi_{p})$ is just accidental. This may be tested by using the spin hydrodynamics which is, however, beyond the scope of this work and we leave it for future.
(4) It is also possible that it is a coincidence from the main assumption
that the spin vector is given by the T-vorticity in the same way as the thermal vorticity.
The true relationship between the spin vector on the freeze-out hyper-surface
and all these vorticities is unclear and has to be figured out.

All our results depend on a set of parameters and assumptions.
For $\mathcal{P}_{z}(\phi_{p})$ and $\mathcal{P}_{y}(\phi_{p})$,
one of the most sensitive parameter is the cutoffs in $p_{T}$ in
Eq. (\ref{eq:pt-int}). For example, as shown in Fig. \ref{fig:pz-ampt-phi},
if we choose the range $p_{T}\in[0,1.2]$ GeV, the theoretical results
match the data of $\mathcal{P}_{z}(\phi_{p})$. But if we choose a
larger range $p_{T}\in[0,3]$ GeV, our theoretical results are much
larger the data of $\mathcal{P}_{z}(\phi_{p})$. The aim of this paper
is a qualitative study instead of a quantitative one. We will carry
out a detailed and quantitative study of the effects in the future.

\section{Summary}

There is a disagreement between theoretical model calculations and
recent experimental data about the azimuthal angle dependence of both
the longitudinal and transverse polarization of hyperons. These theoretical
models are mainly based on the hydrodynamic or kinetic descriptions
of the fluid vorticity and express the spin polarization in terms
of the thermal vorticity. However, away from global equilibrium,
the linear relationship between the spin polarization and thermal vorticity
may not be valid (higher order contribution might be relevant).
Instead, the spin polarization (or equivalently the spin chemical
potential) itself should be regarded as a dynamical variable. Recently
there have been attempts in formulating the theory of relativistic
hydrodynamics with the spin chemical potential as a (quasi-)hydrodynamic
variable, but so far there has been no reliable numerical implementation
of the spin hydrodynamics in the market yet.

In this paper, we assume that the spin vector is determined
from the spin chemical potential $\Omega_{\mu\nu}$ in the same way as
from the thermal vorticity when the thermal vorticity is small,
see Eq. (\ref{average-spin-vector-th}) and (\ref{average-spin-vector}).
We also assume that the spin chemical potential $\Omega_{\mu\nu}$ is still
determined by the fluid velocity and temperature,
which means that $\Omega_{\mu\nu}$ can be regarded as being proportional to a type of vorticity.
In relativistic hydrodynamics there are various types of vorticities such as the kinematic,
temperature and thermal vorticity. There is also a relativistic extension
of the non-relativistic vorticity. We thus explore the possibility
that the spin chemical potential is proportional to these four vorticities
and the spin vector is given by Eq. (\ref{average-spin-vector}).

We use CLVisc, a (3+1)D viscous hydrodynamic model, to compute the
vorticity field. We choose two different initial conditions for the
hydro-simulation: the optical Glauber one without initial orbital
angular momentum and the AMPT one with an initial orbital angular
momentum.
We calculated $\mathcal{P}_{z}(\phi_{p})$ and $\mathcal{P}_{y}(\phi_{p})$
as functions of $\phi_{p}$, the azimuthal angle in transverse momentum,
for four types of vorticities: the kinematic, temperature, thermal
and relativistic extension of the non-relativistic vorticity. Our
results show: (1) All types of vorticities
have the correct sign of $\mathcal{P}_{y}$ for the AMPT initial condition.
With the optical Glauber initial condition, they all give vanishing
results for $\mathcal{P}_{y}(\phi_{p})$ since there is no orbital
angular momentum encoded in the initial state.
For $\mathcal{P}_{y}(\phi_{p})$ with the AMPT initial
condition, only the temperature vorticity has the same trend as the data, 
although its magnitude does not agree with the data.
(2) For the azimuthal angle distribution in the longitudinal
polarization, $\mathcal{P}_{z}(\phi_{p})$, only the temperature
vorticity reproduces the sign of the oscillation
in the azimuthal angle in data. Other three types of vorticities
have a sign difference from the data.
(3) The oscillation behavior of $\mathcal{P}_{z}(\phi_{p})$ (not the magnitude)
is insensitive to the initial conditions with or without
the orbital angular momentum encoded.

\begin{acknowledgments}
The authors thanks F. Becattini, X.L. Sheng and X.L. Xia for insightful discussions.
HZW and QW are supported in part by the National Natural Science Foundation of China (NSFC)
under Grant No. 11535012 and No. 11890713, and the Key Research Program of the Chinese Academy
of Sciences under the Grant No. XDPB09. XGH is supported by NSFC under Grants No. 11535012 and No. 11675041.
\end{acknowledgments}

\bibliography{ref}

\end{document}